\documentclass[trackchanges]{aastex701}
\shorttitle{Random Forest Artifact Identification}
\shortauthors{Einwalter et al.}
\received{December 24, 2025}
\revised{February 19, 2026}
\submitjournal{ApJ}
\begin{document}

\title{DRAGNs in the Forest: Identifying Artifacts with Random Forest Models in the VLASS DRAGNs Catalog}

\author[0009-0002-5763-1199]{Verene Einwalter}
\affiliation{Department of Astronomy, University of Wisconsin-Madison, 
475 N. Charter St., Madison, WI 53703, USA}
\email[show]{verene.einwalter@wisc.edu}

\author[0000-0003-0713-3300]{Eric~J. Hooper}
\affiliation{Department of Astronomy, University of Wisconsin-Madison, 
475 N. Charter St., Madison, WI 53703, USA}
\email[show]{ehooper@astro.wisc.edu}

\author[0000-0001-9920-0210]{Melissa~E. Morris}
\affiliation{Lycoming College Astronomy, One College Place, Williamsport, PA 17701-5192, USA}
\email[show]{morrism@lycoming.edu}

\author[0009-0007-3888-4031]{Sarah Bach}
\affiliation{Lycoming College Astronomy, One College Place, Williamsport, PA 17701-5192, USA}
\email[show]{sarah.bach2736@gmail.com}

\author[0000-0003-1432-253X]{Yjan~A. Gordon}
\affil{Department of Physics, University of Wisconsin-Madison, 
1150 University Ave, Madison, WI 53706, USA}
\email[show]{yjan.a.gordon@gmail.com}

\begin{abstract}
The \textit{Quick Look} data products from the Very Large Array Sky Survey (VLASS) contain widespread imaging artifacts arising from the simplified imaging algorithm used in their production. The catalog of double radio sources associated with active galactic nuclei (DRAGNs) found in the VLASS first epoch \textit{Quick Look} release using the DRAGNhunter algorithm suffers from contamination from these artifacts. These sources contain two or three individual components, each of which can be an artifact. We train random forest models to classify these DRAGNs based on the number of artifacts they contain, ranging from zero to three artifacts. We optimize our models and mitigate the class imbalance of our dataset with judicious training set selection, and the best of our models achieves a weighted F1 score of $97.01\%^{+1.12\%}_{-1.32\%}$. Using our classifications, we produce a catalog of VLASS DRAGNs from which an estimated 99.3\% complete catalog of 97.7\% artifact-free sources can be extracted.

\end{abstract}
\section{Introduction} \label{sec:intro}
One type of object that can be studied with the Karl G. Janksy Very Large Array Sky Survey \citep[VLASS;][]{lacy_karl_2020} is Active Galactic Nuclei (AGN), compact regions at the center of galaxies which emit large amounts of energy and are associated with supermassive black holes. A small fraction of AGN produce visible relativistic jets which interact with the surrounding environment to create plumes of extended emission and sometimes bright hotspots depending on the distance over which they decelerate to nonrelativistic speeds \citep{hardcastle_radio_2020}. These extended sources traditionally appear in a double-lobed structure and hence are known as double radio sources associated with AGNs, or DRAGNs \citep{leahy_dragns_1993}. 

The massive size of VLASS necessitates the usage of automated source-finding methods, however no approach is perfect when using noisy data. The VLASS first epoch \textit{Quick Look} data release suffers from noise issues due to the simple imagining algorithm applied to each image which is intended for rapid image production. As such many images contain artifacts that automated source-finders may misidentify as real sources \citep{lacy_karl_2020, gordon_quick_2021}. 

\citet{gordon_quick_2023} use the double-lobed morphology of DRAGNs to identify extended radio sources in the VLASS first epoch \textit{Quick Look} catalog \citep{gordon_quick_2021} by grouping nearby sources and creating a new catalog which will henceforth be referred to as the VLASS DRAGNs catalog. They identify 17,724 DRAGNs and determine that $\approx 11$\% of these sources are likely to be spurious detections. They identify three primary causes of spurious detections: 1) image artifacts, 2) extended emission broken up into multiple sources, and 3) association of physically unrelated candidate components. The first of these types, artifacts, is the most common cause of spurious detections.

Reducing the human burden of artifact identification is beneficial for the latest radio sky surveys such as VLASS that are detecting tens of thousands of new DRAGNs. For surveys with next-generation radio telescopes such as the SKA, automating artifact identification will be essential. The Evolutionary Map of the Universe survey \citep[EMU;][]{norris_evolutionary_2021, hopkins_evolutionary_2025} uses the Australian Square Kilometre Array Pathfinder telescope \citep[ASKAP;][]{hotan_australian_2021} and is expected to detect hundreds of thousands of new DRAGNs \citep{norris_emu_2025}. The increased sensitivity of the Square Kilometre Array \citep[SKA;][]{dewdney_square_2009} and the Next Generation VLA \citep{murphy_ngvla_2018} will only increase this number further, making visual inspection of all the DRAGNs prohibitively impractical. 

Random forest models, a special type of decision tree models, are a form of supervised learning and a promising method to alleviate this issue. They use an ensemble of decision trees where each tree determines the class of a dataset element by splitting each element's parameters. The resultant final class is determined by combining the predictions of each tree in the ensemble. Random forest models train each decision tree on a random subset of the parameters in the training set, thus mitigating issues with overfitting which are common with traditional decision trees \citep{ho_random_1998, breiman_random_2001}. Compared to other supervised classification methods like Naive Bayes, k-nearest neighbors, and support vector machines, random forest models provide better performance and are an attractive option for use in astronomy \citep{fernandez-delgado_we_2014, solorio-ramirez_random_2023}. 

Random forest models have been used to classify a range of astronomical objects, including AGN \citep[e.g.][]{de_cicco_selection_2025}, white dwarfs \citep[e.g.][]{garcia-zamora_random_2025}, supernovae \citep[e.g.][]{jayasinghe_asas-sn_2018}, and stellar spectral types \citep[e.g.][]{li_stellar_2019}. These models have also been employed to differentiate between astronomical objects like stars and galaxies using redshift and multi-wavelength survey magnitudes \citep[e.g.][]{asadi_semi-supervised_2025} as well as estimating photometric redshift probability density functions \citep[e.g.][]{carrasco_kind_tpz_2013}. \citet{helfand_last_2015} use an ensemble of oblique decision trees, similar to a random forest model, in Faint Images of the Radio Sky at Twenty-centimeters \citep[FIRST;][]{becker_first_1995} to determine the probability that a given source is a sidelobe. As such, there is precedence for their efficacy in astronomy applications. 

If trained on a sufficiently-sized dataset, a random forest model can reach high levels of accuracy and reliably classify a much larger dataset \citep{ho_random_1998}. We can eliminate the bulk of spurious detections caused by imaging artifacts in VLASS \textit{Quick Look} series by training a random forest model to identify sources with artifacts in the DRAGNs catalog. For the comparatively small cost of manually identifying artifacts  in a small subset of sources, a random forest model trained to identify artifacts can significantly improve the accuracy of published catalogs by classifying sources based on the number of artifacts they contain. A catalog utilizing this method can improve accuracy by filtering out artifact-containing sources to create a cleaner catalog. 

\citet{vantyghem_rotation_2024} use a self-organizing map, an unsupervised clustering algorithm, to identify the probability that a single source in the VLASS \textit{Quick Look} first epoch catalog is a sidelobe. Though we focus on multi-component sources in the same catalog, our random forest model has the advantages of being less computationally intensive and providing additional information by discriminating between the number of artifacts in a multi-component source. 

In section \ref{sec:motivation} we motivate our utilization of a random forest model by identifying preliminary correlations between parameters in the VLASS DRAGNs catalog and the number of artifacts in each source. Following, in section \ref{sec:methods} we describe how we prepared our data for classification, and discuss our methods for selecting training sets and finding the best parameters for our random forest models. In section \ref{sec:doubles_class} we present our two best models, one trained on the triple-component subset of DRAGNs, and the other trained on a specially-selected set of double-component DRAGNs, and we evaluate their performance classifying a representative sample of double-component DRAGNs. We also compare the performance of our random forest model to the existing method of filtering spurious sources in the VLASS DRAGNs catalog. In section \ref{sec:discussion} we discuss the limitations and other details of our random forest models. Finally, in section \ref{sec:conclusion}, we summarize our results and our final catalog of DRAGNs with the number of artifacts as identified by the best of our random forest models, and we discuss further applications of this work.
%
%==================================================================================%
%==================================================================================%
%
%
%
%
%==================================================================================%
%==================================================================================%
%
\section{Motivation} \label{sec:motivation}
\subsection{Artifact Visual Classification}\label{subsec:motiv_classification}
The first epoch VLASS \textit{Quick Look} DRAGNs catalog can be separated into two subsets: double radio sources (doubles; 15,888 in total) and triple radio sources (triples; 1,836 in total). These multi-component sources were identified using DRAGNhunter, an automated source-finding algorithm that correlates nearby regions of radio emission into sets of components as described in \citet{gordon_quick_2023}. The doubles are comprised of two components, while triples include a core in addition to two components. The triples set is a significant but tractable fraction of the DRAGNs catalog, and previous, less rigorous, visual inspection of this set has revealed that $\approx13\%$ of triples contain contaminants \citep{gordon_quick_2023}.

We performed visual inspection of the triples set to determine the number of artifacts in each source, ranging from 0 to 3. The number of artifacts corresponds to the artifact class of a source, i.e. 0-, 1-, 2-, and 3-artifact.  For the purposes of our classification, an artifact is any component identified by DRAGNhunter that does not contain real and clearly defined radio emission. We developed a series of qualitative criteria, described below, which we used to evaluate whether each component is an artifact or is associated with real emission, i.e., a real component. To help distinguish between real emission and artifacts we compared each VLASS image to the corresponding region of the sky in the Wide-field Infrared Survey Explorer \citep[WISE;][]{wright_wide-field_2010} using images from the unWISE dataset \citep{lang_unwise_2014, meisner_deep_2017, meisner_full-depth_2017}, which were retrieved from the\citet{unwise_data}.

\begin{figure*}
    \gridline{
        \fig{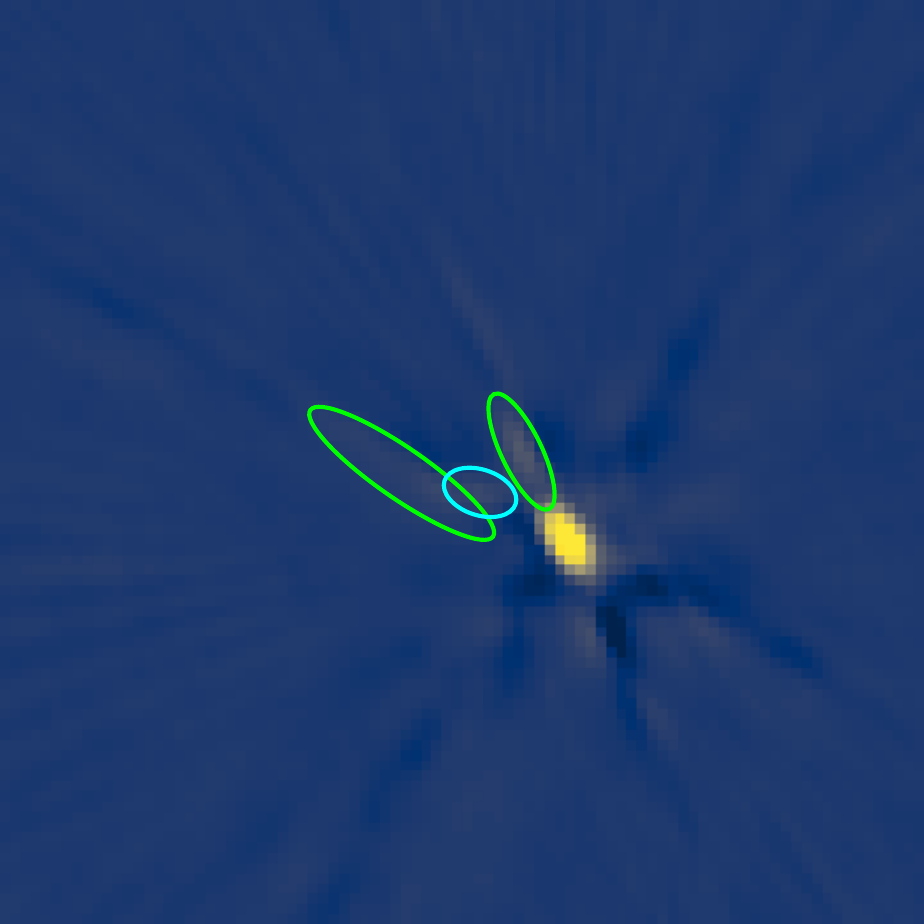}{0.33\textwidth}{(a) J225504.95-084356.0}
        \fig{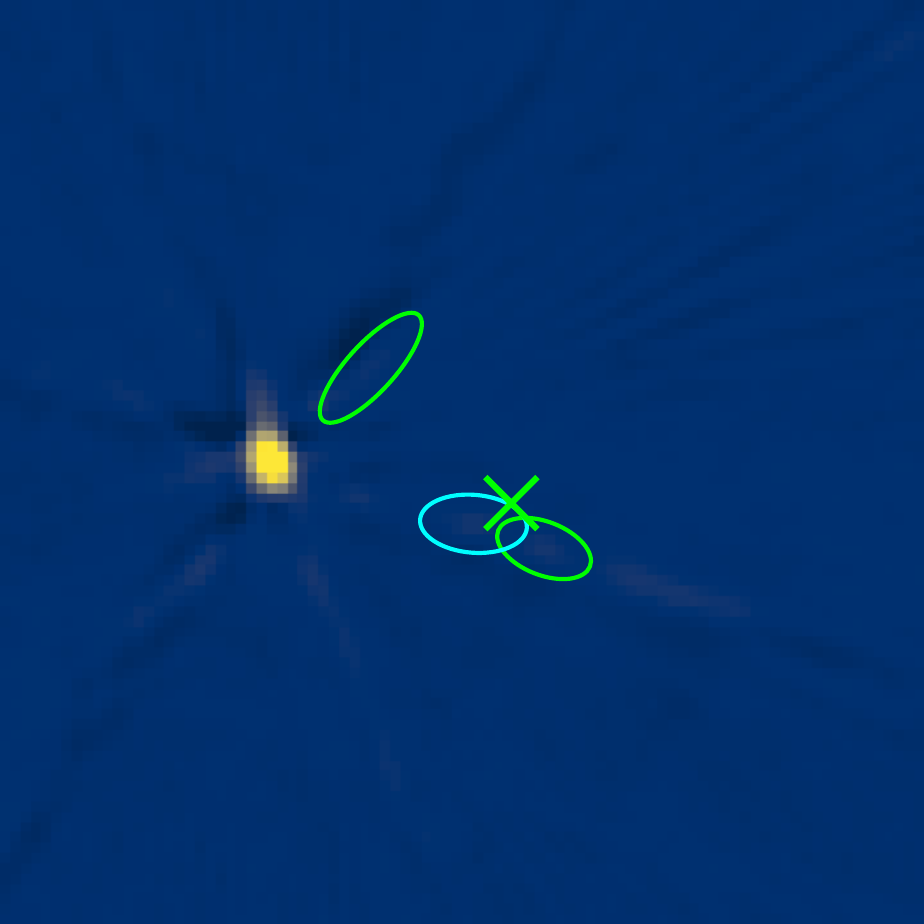}{0.33\textwidth}{(b) J235024.14-022443.2}
        \fig{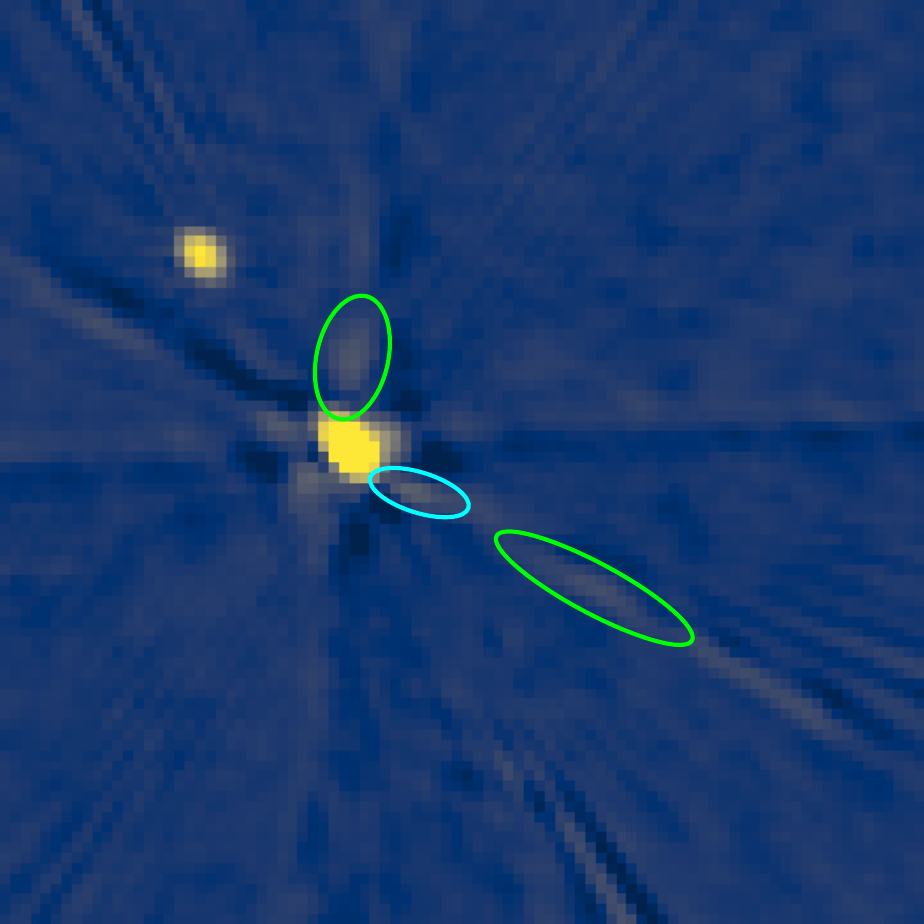}{0.33\textwidth}{(c) J235834.30+440436.8}
        }
    \gridline{
        \fig{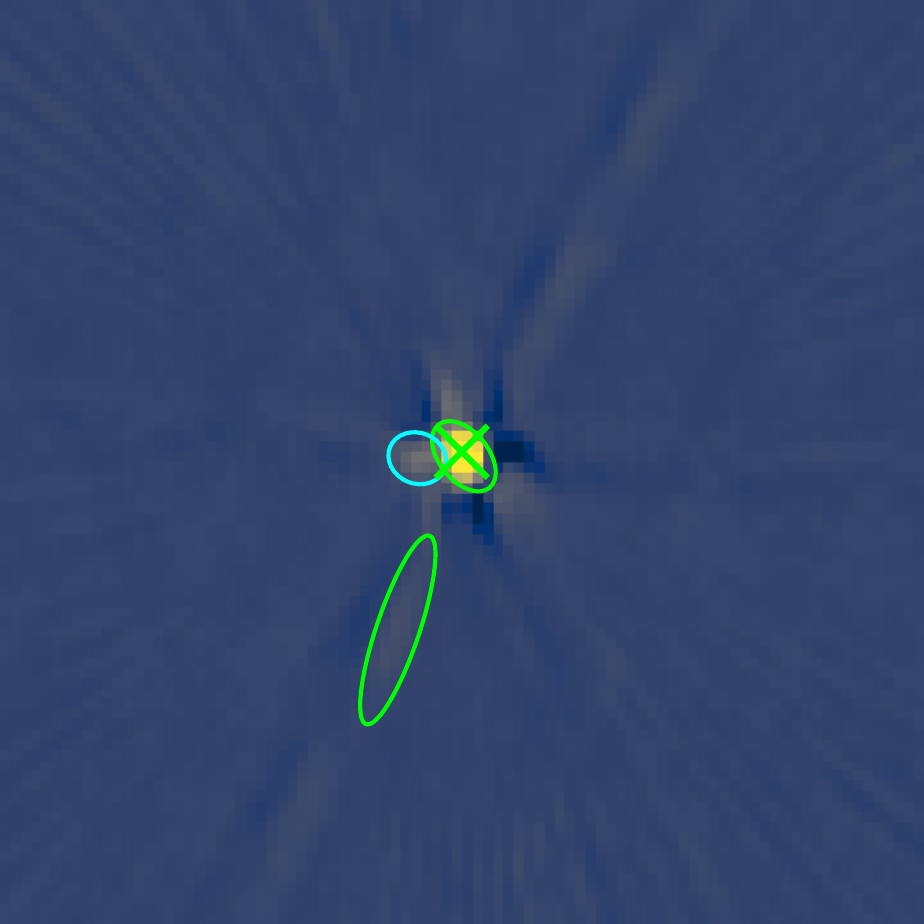}{0.33\textwidth}{(d) J235656.44+675137.7}
        \fig{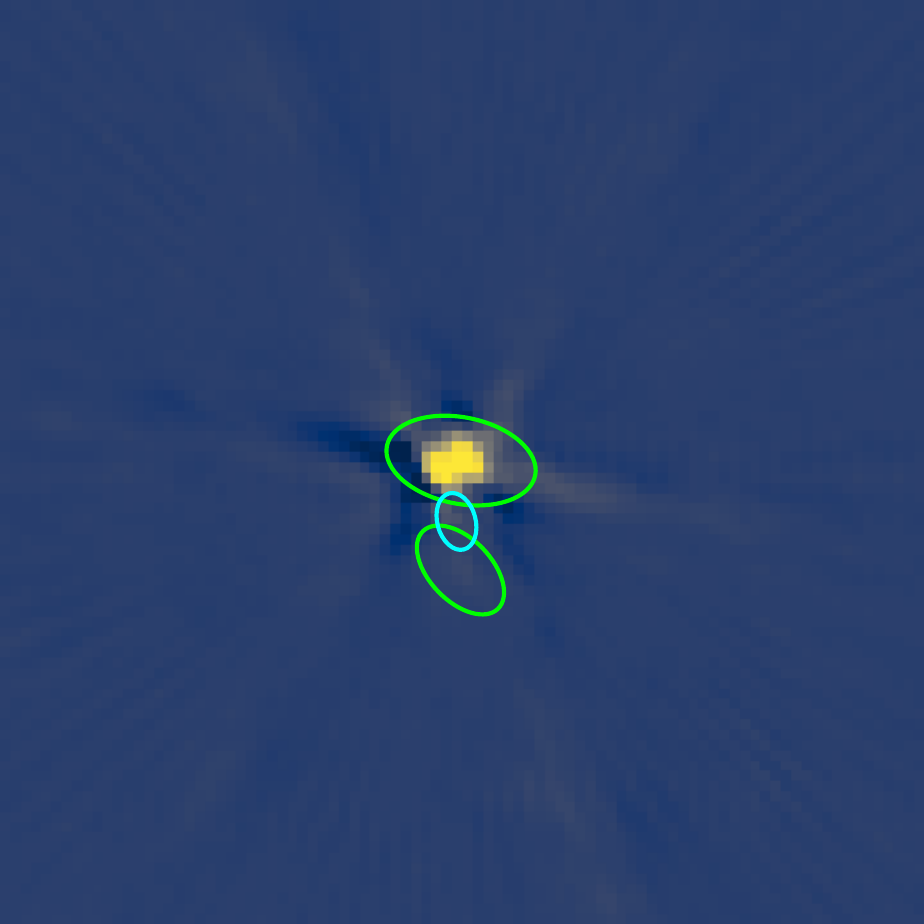}{0.33\textwidth}{(e) J234023.10+222055.0}
        \fig{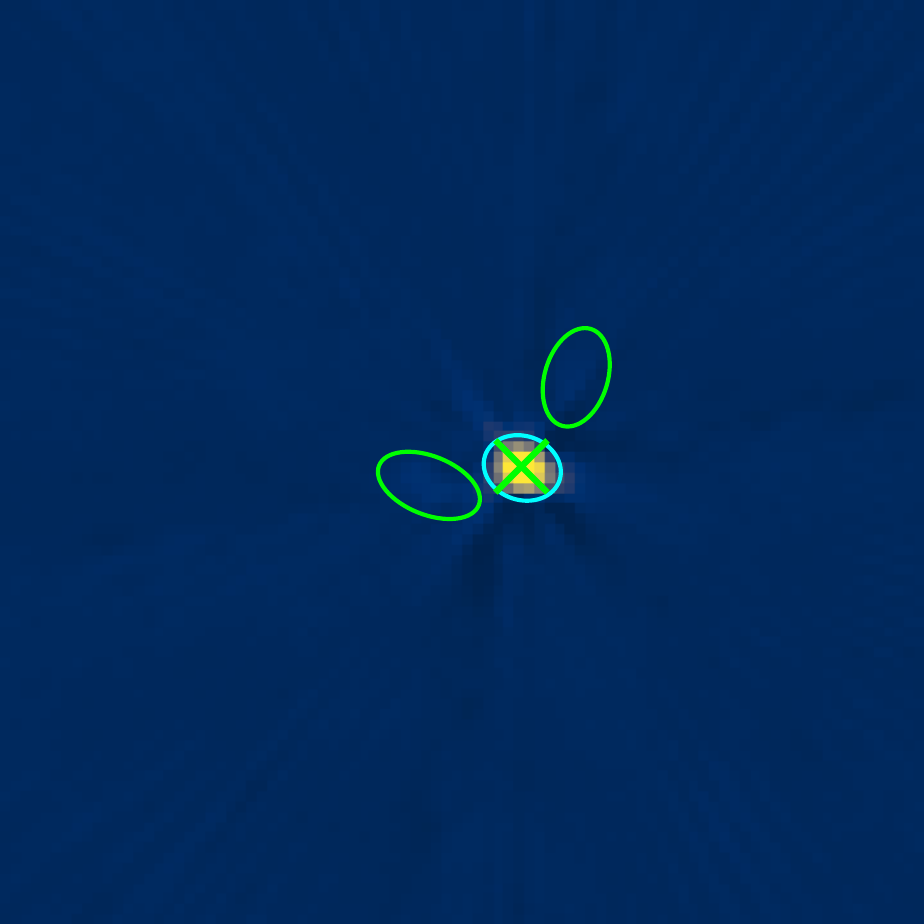}{0.33\textwidth}{(f) J232226.58+505752.4}
        }
    \gridline{
        \fig{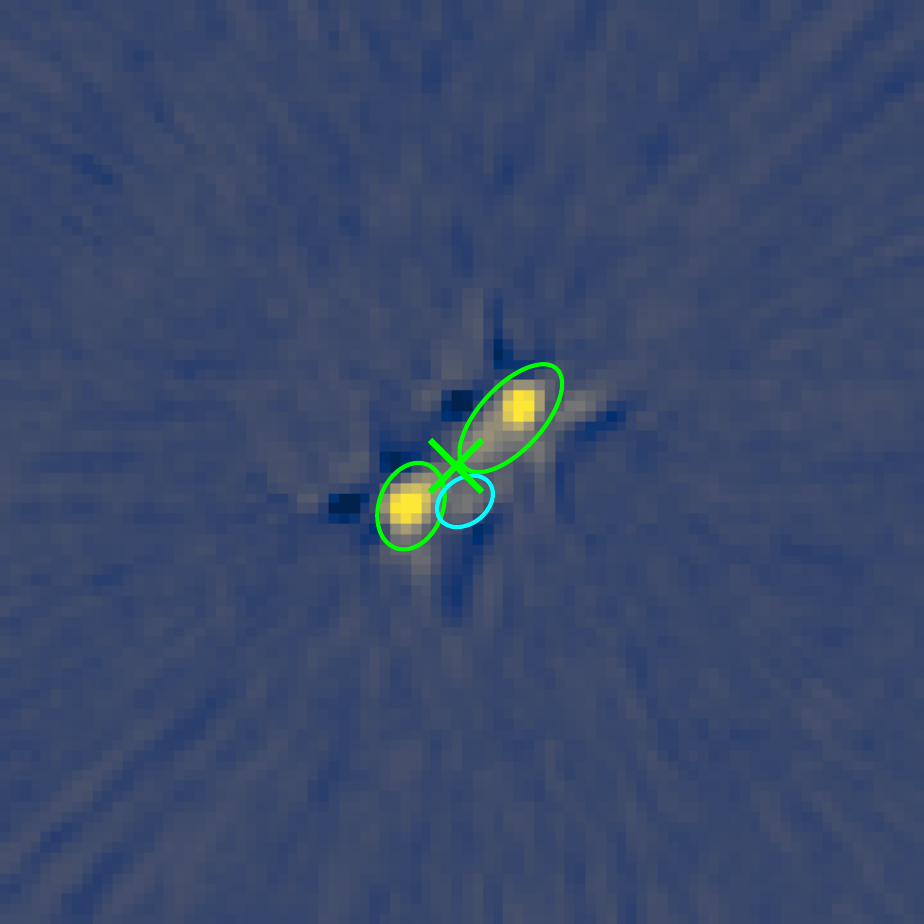}{0.33\textwidth}{(g) J003951.97+045531.7}
        \fig{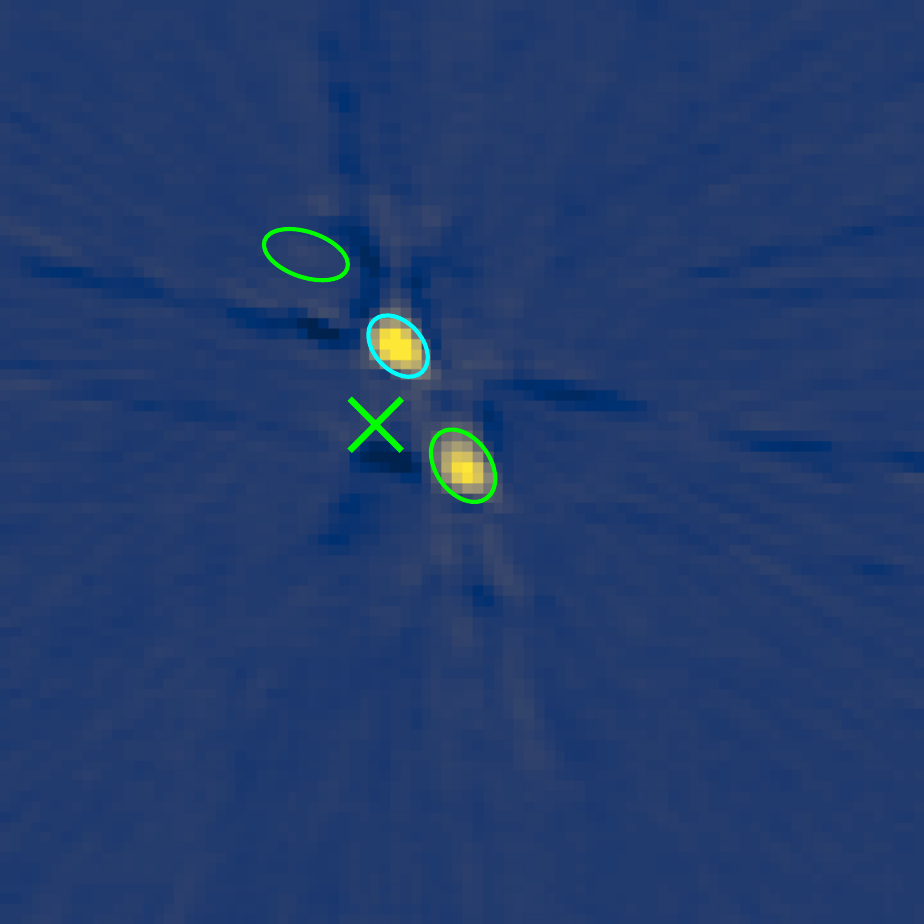}{0.33\textwidth}{(h) J085300.03-204733.2}
        \fig{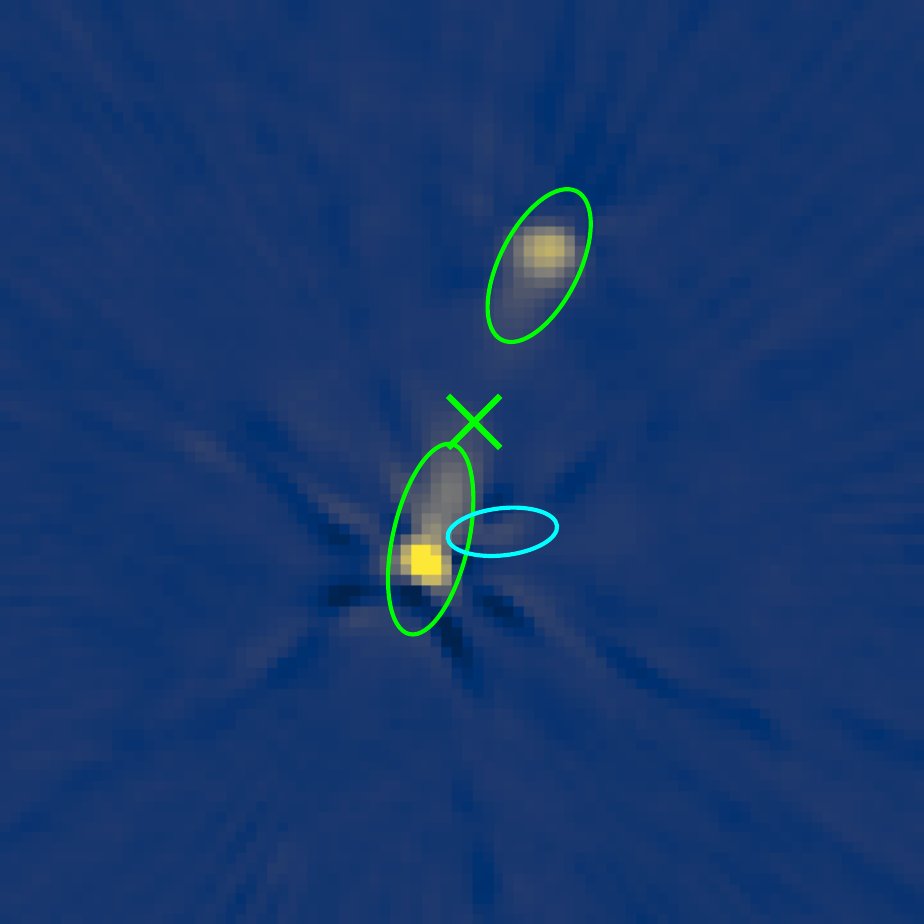}{0.33\textwidth}{(i) J234829.62+184404.9}
        }
    \caption{Collage of 1.5'x1.5' VLASS images of triple-component DRAGNs identified by DRAGNhunter. The ellipses denote components as identified by DRAGNhunter, where the green ellipses denote the lobe or jet hot spot component, the cyan ellipse denotes the identified core, and the green X denotes the AllWISE host as identified in \citet{gordon_quick_2023}, if one was found. \\
    (Top line) Examples of what typical DRAGN triples with 3 artifacts look like. Note that they tend to result from artifacts around very bright point sources (e.g. a, b), which are usually single sources, but can also be very bright jet lobes from extended sources (e.g. c). \\ 
    (Center Line) Examples of what typical 2-artifact triples look like. These are predominantly characterized as 2 artifacts surrounding an unresolved point source. \\
    (Bottom Line) Example 1-artifact triple sources. These are usually double sources with a spurious artifact component around a bright jet lobe or hot spot.}
    \label{fig:triples_a_examples}
\end{figure*}

\begin{figure*}[htb!]
    \gridline{
        \fig{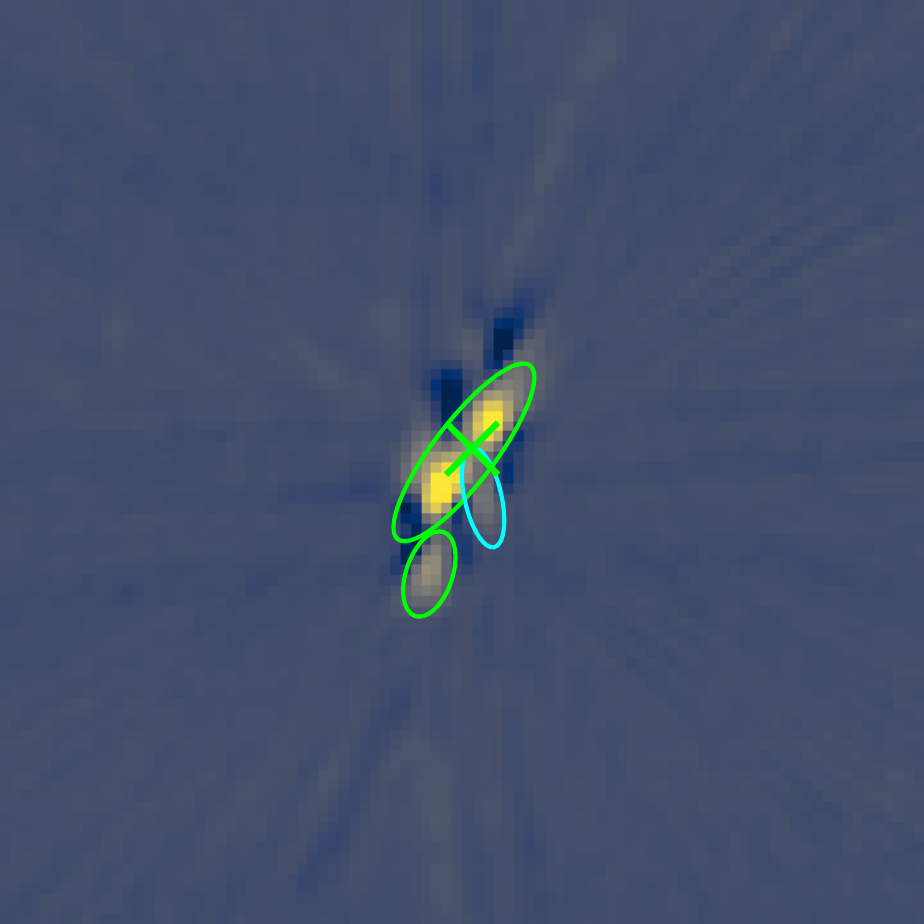}{.25\textwidth}{(a) J024447.90-352305.0 \\ (2 artifacts)}
        \fig{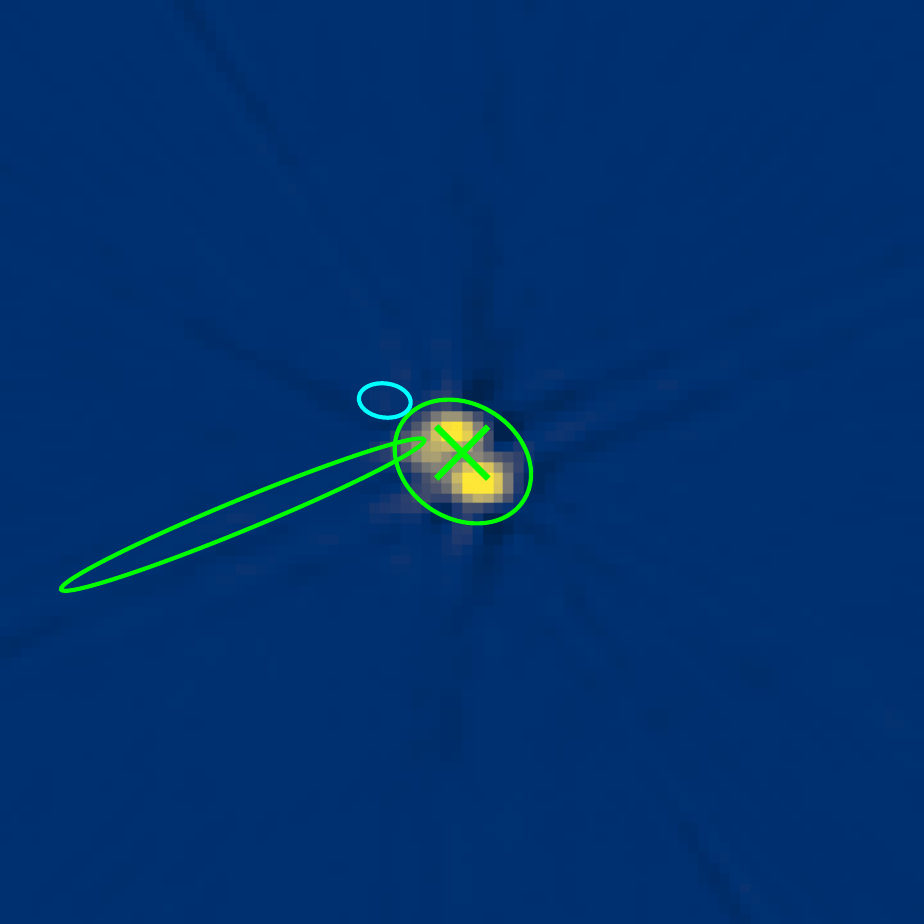}{.25\textwidth}{(b) J081336.05+481301.7 \\ (2 artifacts)}
        \fig{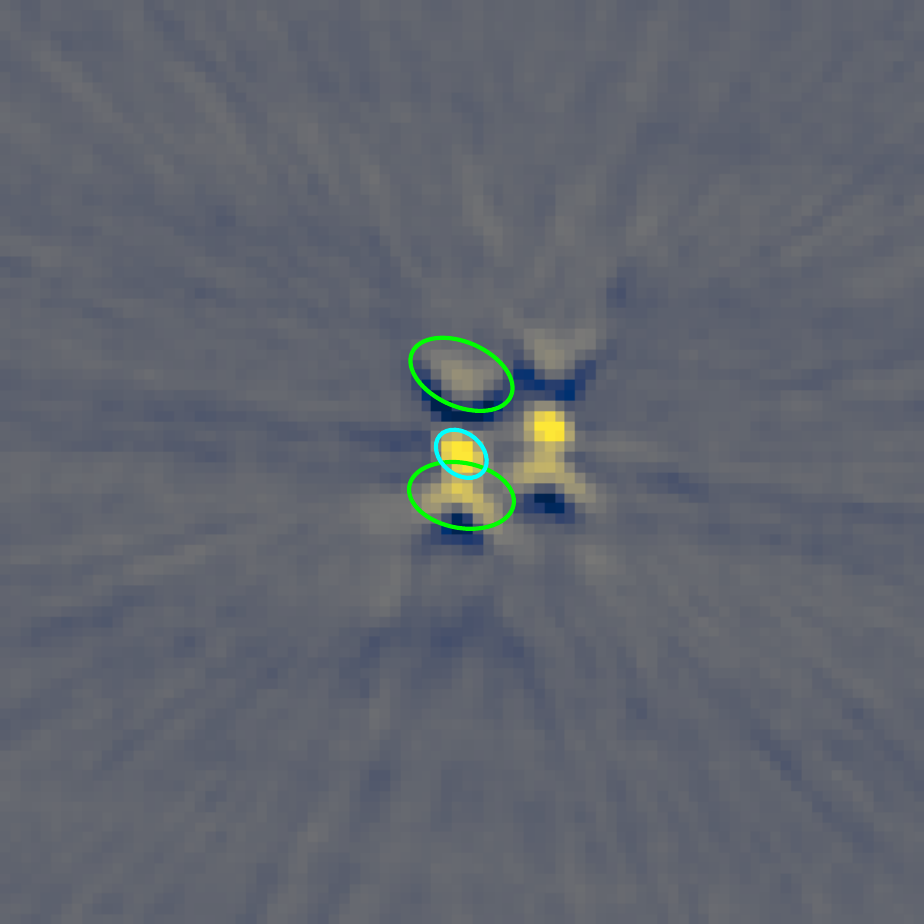}{.25\textwidth}{(c) J093049.12-180930.0 \\ (2 artifacts)}
        \fig{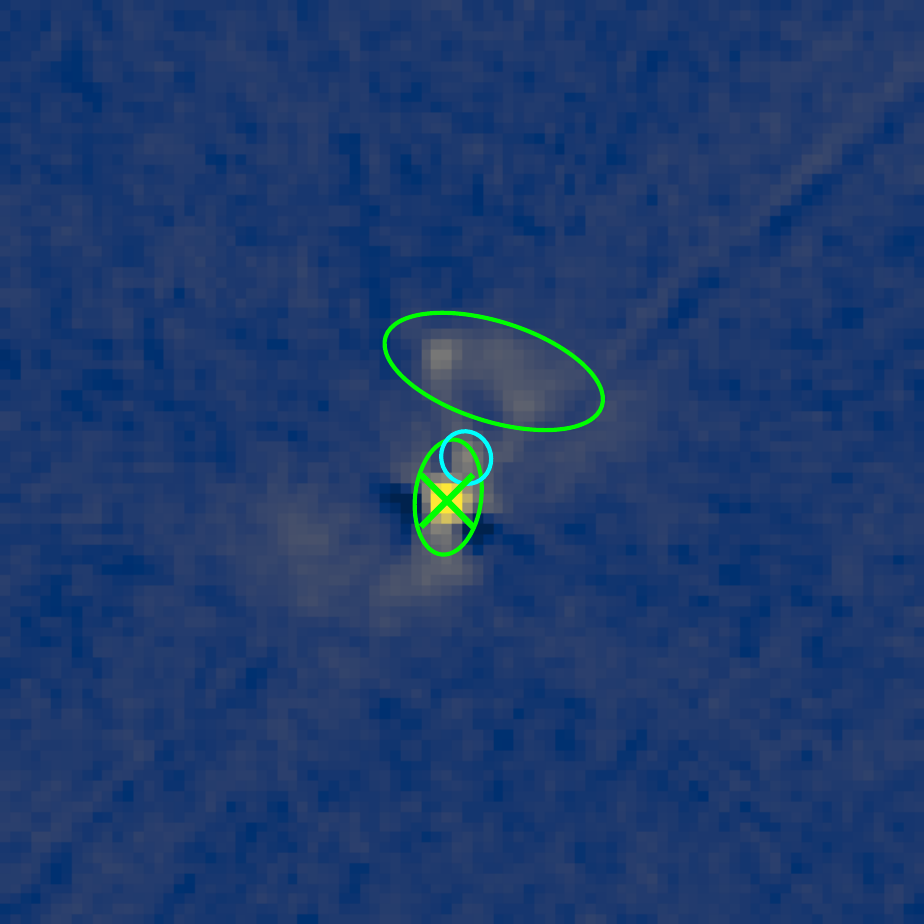}{.25\textwidth}{(d) J111426.99+263302.6 \\ (0 artifacts)}
        }
    \caption{Examples of triple sources with dubious morphologies which are difficult to classify. The ellipses denote components as identified by DRAGNhunter, where the green ellipses denote the lobe or jet hot spot component, the cyan ellipse denotes the identified core, and the green X denotes the AllWISE host as identified in \citet{gordon_quick_2023}, if one was found. (a) A double with a strong Y-shaped sidelobe pattern where the core and bottom component are artifacts. (b) A source where part of one highly elongated component intersects, but does not encompass real emission. Components like this, which are centered on an artifact and happen to overlap with real emission, were classified as artifacts. (c) A source with prominent sidelobes mirrored on both sides of the source; we called these artifacts `tabs'. (d) The diffuse, extended emission of this AGN is faint and difficult to distinguish from background noise, however all components are associated with real, separate emission.}
    \label{fig:triples_dubious}
\end{figure*}

We also used images from data releases 1 and 2 \citep[DR1 and DR2, respectively;][]{panstarrs_dr1, panstarrs_dr2} of the Panoramic Survey Telescope and Rapid Response System survey \citep[PanSTARRS;][]{chambers_pan-starrs1_2016} to help disambiguate source appearance. In each image we also compared our visual impression of potential WISE host galaxies to the host galaxy candidate identified in \citet{gordon_quick_2023} derived from the AllWISE catalog \citep{cutri_explanatory_2012} which is available at \citet{allwise_data}. At least two people inspected every source independently, and if there was disagreement in the artifact class of a source we convened and came to a consensus about its class. We provide example images of triple artifact-containing classes (i.e. 1, 2, or 3 artifacts) in Fig. \ref{fig:triples_a_examples}, as well as images of difficult-to-classify sources in Fig. \ref{fig:triples_dubious}.

We used a series of qualitative tests to determine whether a given component is an artifact. Each component was evaluated based on the characteristics of the emission it's centered upon and best encompasses. Most real components are centered on and encompass clear emission above the noise floor, and thus they are readily apparent in the majority of cases. For dimmer components, or those which are complicated by nearby sidelobes, we can use other morphological patterns to distinguish them from artifacts. Coincidence with a potential host galaxy in the corresponding PanSTARRS or WISE images strongly implies that the component is real. Additionally, components associated with jet lobes are real and are often readily apparent, as they almost always are paired with a lobe from the opposing jet of the AGN host. Jet lobes are further distinguished by identifying whether they are continuous with jet emission, or are elongated with a more faint boundary towards the assumed radio core, such as in Fig. \ref{fig:triples_a_examples}(i). Components that are associated with jet hotspots and trailing plumes are morphologically distinct and are similarly real. Finally, components centered on extended diffuse emission that aren't coincident with a host galaxy or are associated with a nearby AGN, but do not satisfy any artifact criteria were identified as real. This includes extended diffuse emission associated with the plane of the Milky Way. 

Components that did not satisfy any of the criteria of real components were then evaluated based on our artifact criteria. A component that is aligned with the sidelobe pattern originating from a nearby bright source is highly suspect, and these components are often elongated such as in Fig. \ref{fig:triples_a_examples}(a,c,d). Some artifacts are apparent sources that cluster tightly around a bright point source, with peaks that are along the sidelobe pattern. These artifacts, which we called `tabs', have a mirror on the other side of the source corresponding to an area with apparent negative flux. These tabs appear in most images with bright emission, such as the short, mirrored v-shaped spikes around the lobes in Fig. \ref{fig:triples_dubious}(c). Components that satisfied one or more of these criteria, and did not satisfy any criteria of real components, were identified as artifacts.

In a minority of sources there are overlapping components, such as in Fig. \ref{fig:triples_dubious}(d). Overlapping components suggest that there are multiple sources of emission in a concentrated area, such as a region of diffuse emission overlapping with a jet hotspot, or a sidelobe intersecting with a DRAGN core. In these cases we identified the most likely source of the emission that the component encompasses, i.e., a jet lobe, core, hotspot, area of extended emission, or sidelobe. Components which encompass emission that is most likely caused by a sidelobe or other artifact are spurious. For example, Fig. \ref{fig:triples_dubious}(d) shows a source with two overlapping components, yet both encompass real emission—one centered on the AGN core and the other on a hotspot.

Some more peculiar sources may be excluded through these criteria, such as hotspots or areas of diffuse emission which are coincident with very strong artifacts. However, this possibility is minimal, as we enforced a stronger burden of evidence for a component to be classified as an artifact than a classification of real so as to avoid bias against unusual sources. Fig. \ref{fig:triples_dubious}(d) provides an example of this, where the cyan core component is coincident with both a sidelobe and what appears to be a jet hotspot, and the former evidence is not sufficient for us to say with certainty that it is an artifact. We carefully evaluated these edge cases and were generally conservative in labeling them as artifacts. We found that $15.2\%$ of triple sources contain at least 1 artifact. The population of each artifact class is provided in Table \ref{tab:triples_class_fraction}. 

\begin{deluxetable}{c c c c c}[htb!]
    \tablecaption{Triples artifact class fractions as identified by visual inspection}
    \label{tab:triples_class_fraction}
    \tablehead{
        \colhead{Number of Artifacts} & \colhead{0} & \colhead{1} & \colhead{2} & \colhead{3}
    }    
    \startdata
        Number of Sources & 1557 & 37 & 200 & 42 \\
        Percent of Total & 84.8 & 2.0 & 10.9 & 2.3
    \enddata
\end{deluxetable}

While most sources are straightforward to classify like those in Fig. \ref{fig:triples_a_examples}, other sources are more dubious. The primary issue is that around some sources, particularly bright, unresolved point sources and jet features, the noise floor is elevated, and it is difficult to distinguish real, diffuse emission from artifacts, such as the sources in Fig. \ref{fig:triples_dubious}(a,c). Though we came to a consensus on all sources, sources that straddle the division between artifact classes introduce some inherent uncertainty in our observations. 

\begin{figure}[htb!]
    \epsscale{.6}
    \plotone{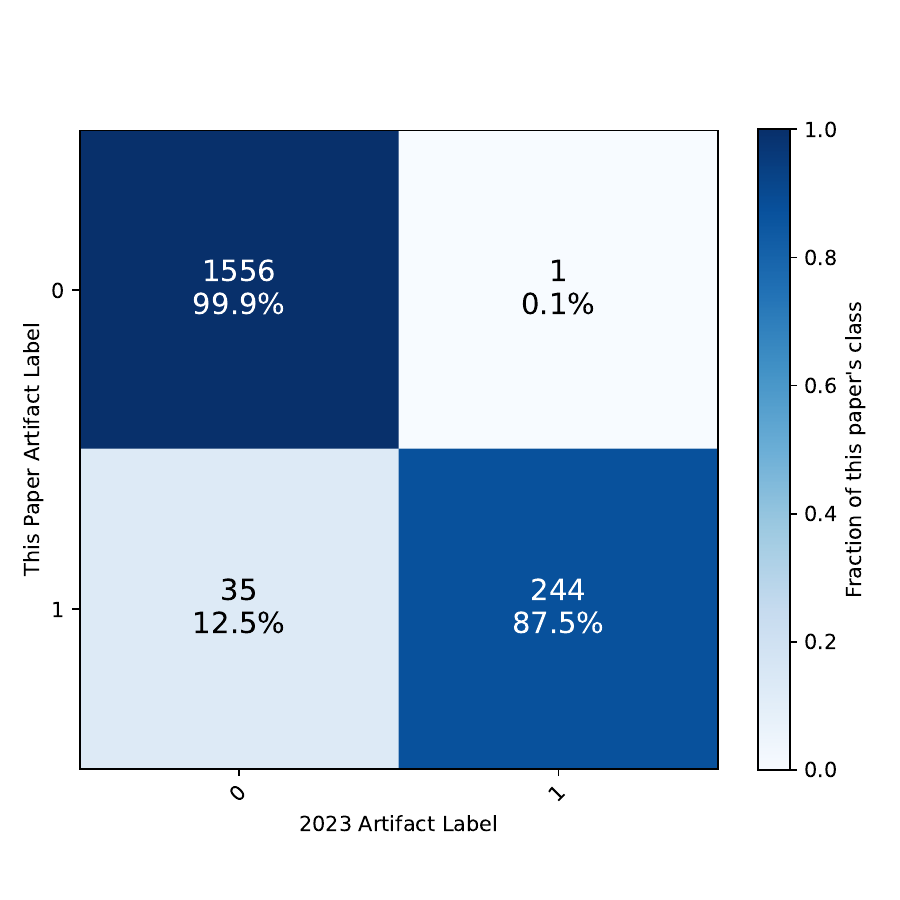}
    \caption{Confusion matrix comparing the results of visual inspection for artifacts of this paper and spurious detections of the triples in \citet{gordon_quick_2023}, where 1 denotes spurious and 0 denotes not spurious. Spurious sources are those that contain artifacts. The percentages and the color of each square are determined by the fraction of the total population of the row, i.e. the fraction of the true class, that is present in each square. All sources that we identified in this paper as having 1 or more artifacts were labeled as 1 in this matrix.}
    \label{fig:spurious_matrix}
\end{figure}

We compare the results of this paper's visual artifact classification to visual inspection in \citet{gordon_quick_2023} in Fig. \ref{fig:spurious_matrix} using a confusion matrix. These graphs compare how each visual inspection pass classified sources, where agreement appears as high population fractions along the diagonal, and disagreement elsewhere. We identify 2\% more sources as having artifacts in this paper than in the preceding visual inspection. This discrepancy is in part due to the increased complexity of the visual inspection employed in this paper, which was more intensive than the binary real or spurious classification employed previously. Counting the number of artifacts and comparing the location of sources to potential WISE and PANSTARRs hosts introduced additional checks that brought more attention to whether particular components are or are not artifacts. Additionally, the team performing artifact identification in this work is larger than the team in \citet{gordon_quick_2023}, and more people independently inspected each individual source, thus improving the reliability of our classifications. Thus, we conclude that our artifact class inspection is more complete.

%
%
%==================================================================================%
\subsection{Preliminary Clustering} \label{subsec:prelim_clustering}

As an important initial step, we analyzed the parameters derived by DRAGNhunter in the triples catalog with simple plots to identify possible clustering between different artifact-count subsets. The DRAGNs catalog contains measurements of the largest angular size (LAS) and flux of each source, in addition to component flux ratios, core prominence, and other parameters, which can be used to analyze each source. Following the recommendation of \citet{gordon_quick_2021} we apply a correction and multiply the flux values by $1/0.87$ to account for systemic underestimation of flux values in the VLASS first epoch \textit{Quick Look} data release. 

 Prominence is a useful analysis parameter because it characterizes how bright a particular source component is compared to the total brightness of a source. Though the prominence of the core is provided in the DRAGNs catalog, we calculated and used the prominence of the brightest component to analyze the sources in the DRAGNs catalog. For the majority of triples, we expect that the prominence of the core will be low because the core is usually much dimmer than the jet lobes. In the largest artifact-containing class, 2-artifact, we usually see that one component is associated with an unresolved point source while the other components are associated with noisy space around the point source and/or sidelobes of a single source. These other regions contain much less flux than the point source. Hence, the core prominence of 2-artifact sources should be much higher than that of regular triples, however the identified core component is not always the component associated with the actual point source. Hence, we calculate the prominence of brightest component using the fraction of total flux contained in the flux of the brightest component. 

\begin{figure*}[htb!]
    \epsscale{1.0}
    \gridline{
    \fig{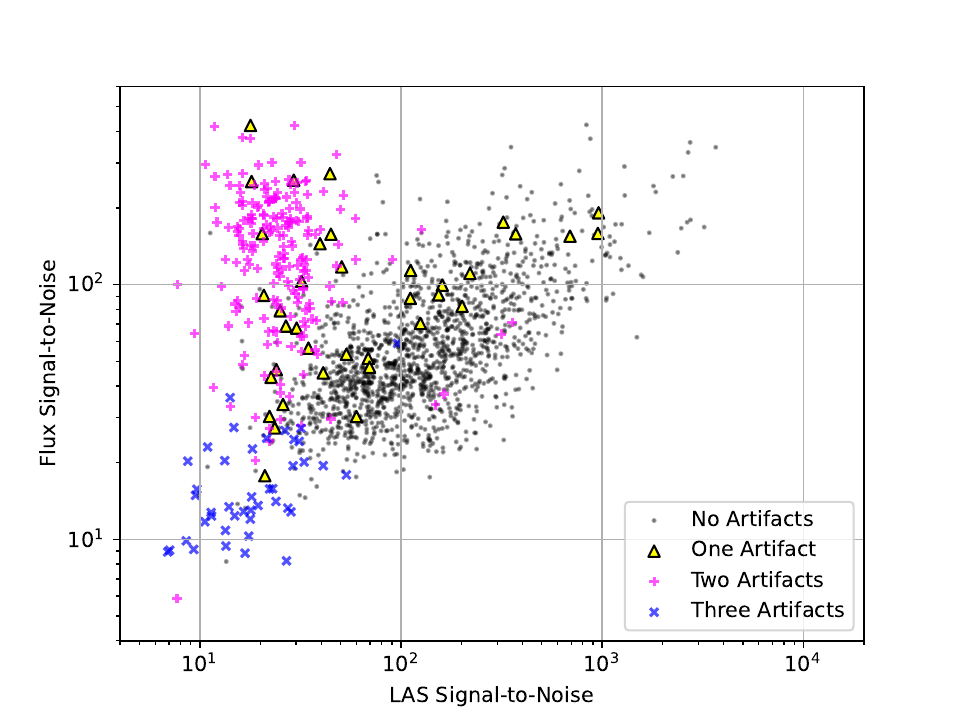}{0.5\textwidth}{(a) Triples}
    \fig{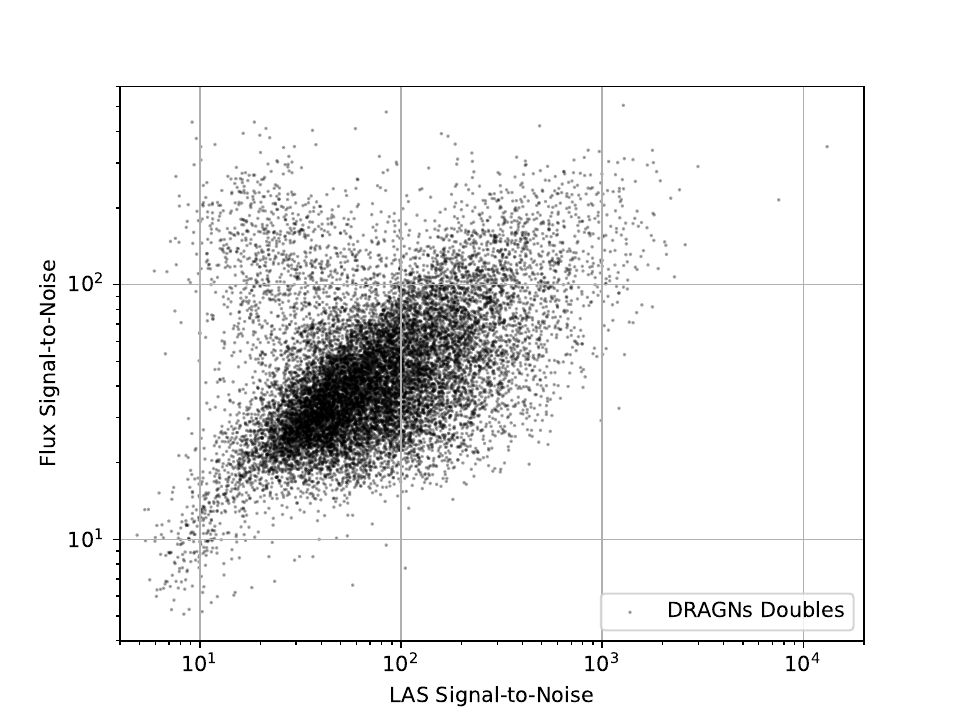}{0.5\textwidth}{(b) Doubles}
    }
    \caption{(a) Scatterplot of LAS S/N vs. Flux S/N for all triples grouped by number of artifacts in each source as identified by visual inspection. This particular set of parameters shows that the 0-, 2-, and 3-artifact classes of sources cluster in approximately 3 separate areas. (b) Scatterplot of LAS S/N vs. Flux S/N for all doubles which suggests that the double sources may also cluster into artifact classes within this parameter space like the triples.}
    \label{fig:las_flux_sn}
\end{figure*}

\citet{gordon_quick_2023} identify the signal-to-noise (S/N) of the LAS as a key parameter for separating discrepant sources from real sources. Using this parameter, we can extend the separation of artifact classes if we look at another parameter derived from the DRAGNhunter catalog, the S/N ratio of the total flux. We plot the S/N of the LAS and S/N of the flux of each triple source in Fig. \ref{fig:las_flux_sn}(a). Not only do we see clear separation between 0-artifact and 2-artifact sources, we also see separation and clustering of 3-artifact sources. We plot the same parameters for the doubles sources in Fig. \ref{fig:las_flux_sn}(b) for comparison between the source subsets.

This preliminary plot shows that there is correlation between the catalog parameters and the number of artifacts in each source of the DRAGNs catalog. While we could use traditional 2-dimensional clustering methods like k-means clustering to find these clusters in flux S/N and LAS S/N space, this does not take advantage of the full range of parameters available to us in the catalog. A more scalable solution, which could be applied to more than just the original dataset, would be to train a random forest model to classify these triple sources by number of artifacts. If applied to a second dataset with the same parameters as the training dataset, and similar feature distribution, it would be possible to bypass the arduous human visual classification step and find artifacts in much larger samples of sources. The VLASS DRAGNs catalog presents a prime opportunity for this, as there are 15,888 DRAGNs doubles which would be prohibitively time-consuming to visually classify manually. 

%
%
%==================================================================================%
\subsection{Analogy Between Triple and Double Artifact Classes} \label{subsec:motivation_analogy}

Ideally, one could train a random forest model on the triples and apply it to determine each double's artifact class. It is important, however, to evaluate whether double sources are similar enough to triple sources for a random forest algorithm trained on triples to be applicable without a significant loss in accuracy. This requires two things to be true: 1) the artifact classes in the triples set are analogous to those in the doubles set, and 2) artifact classes in the doubles set cluster in the same parameter space and range as artifact classes in the triples set. 

\begin{figure*}[htb!]
    \gridline{
        \fig{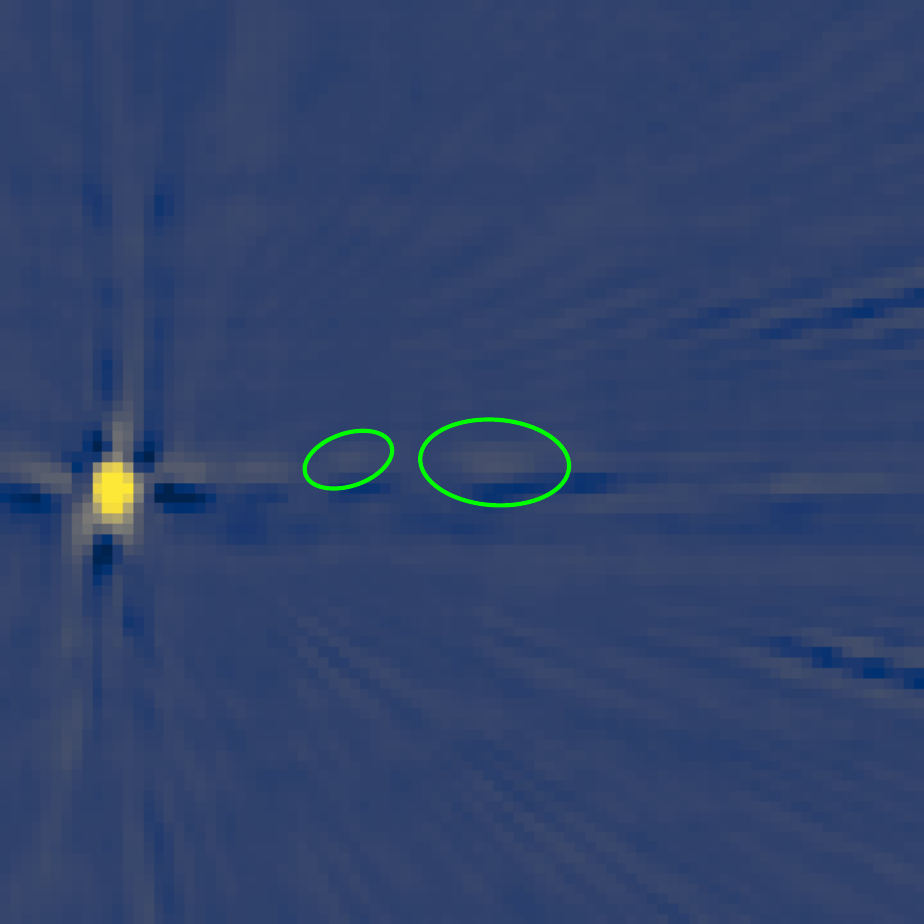}{0.33\textwidth}{(a) J022253.62-344125.7}
        \fig{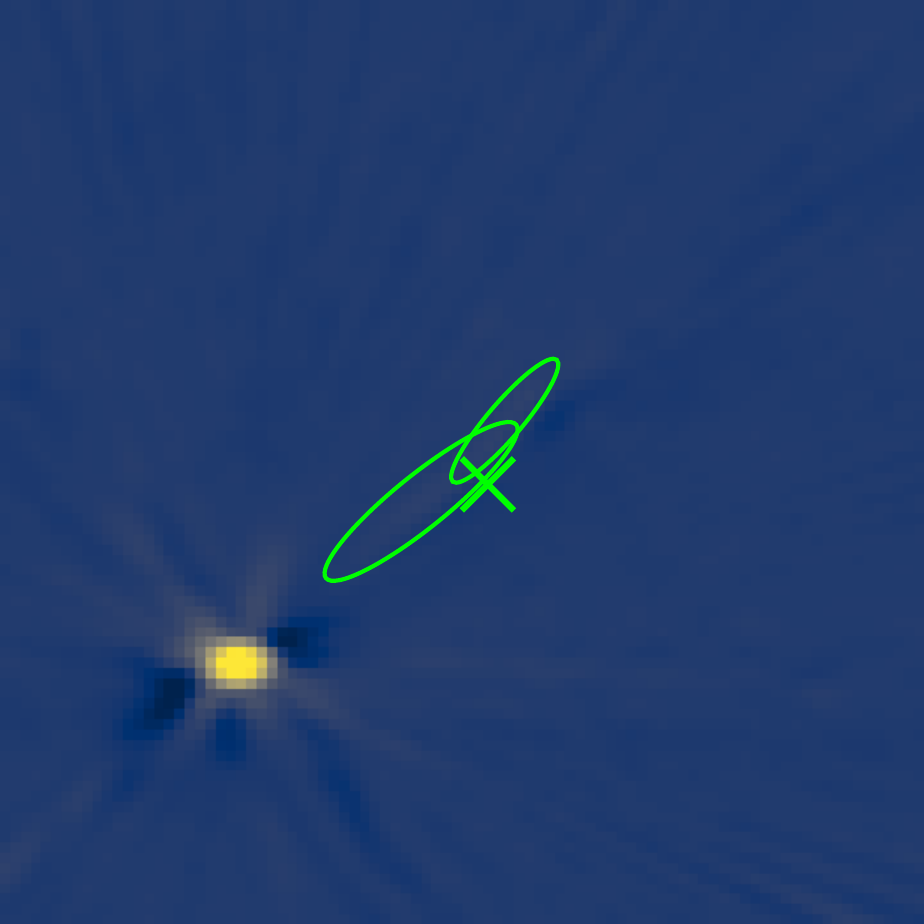}{0.33\textwidth}{(b) J143841.60+621214.0}
        \fig{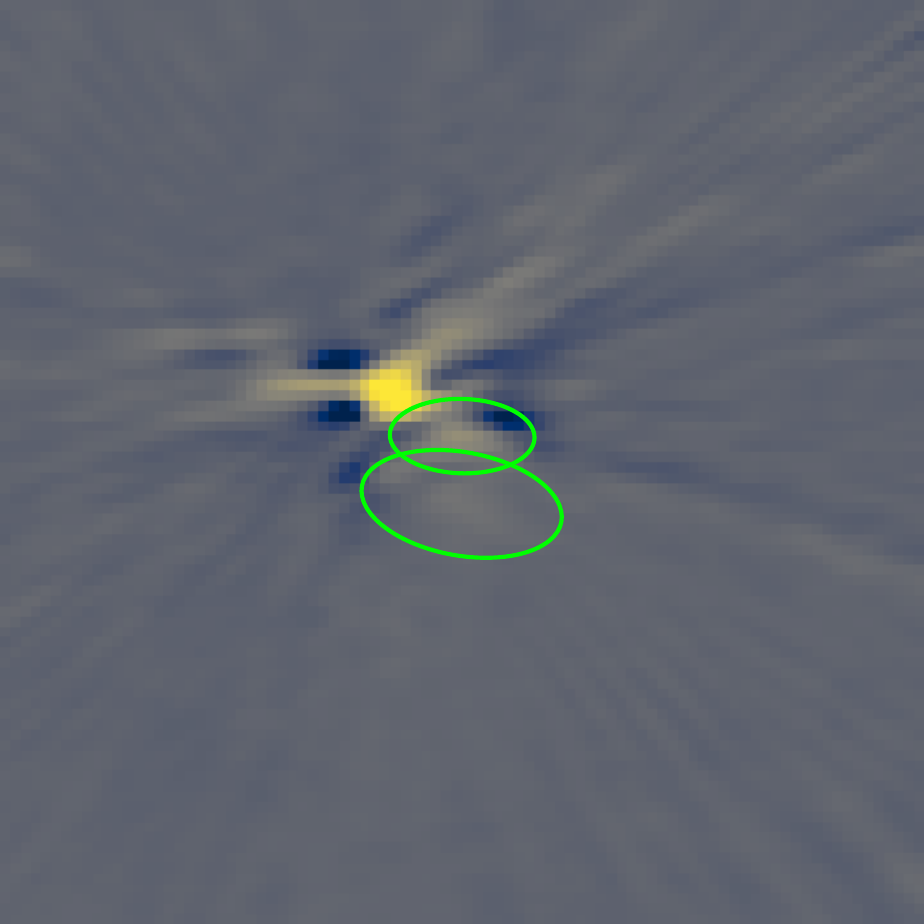}{0.33\textwidth}{(c) J235619.24+815245.7}
        }
    \gridline{
        \fig{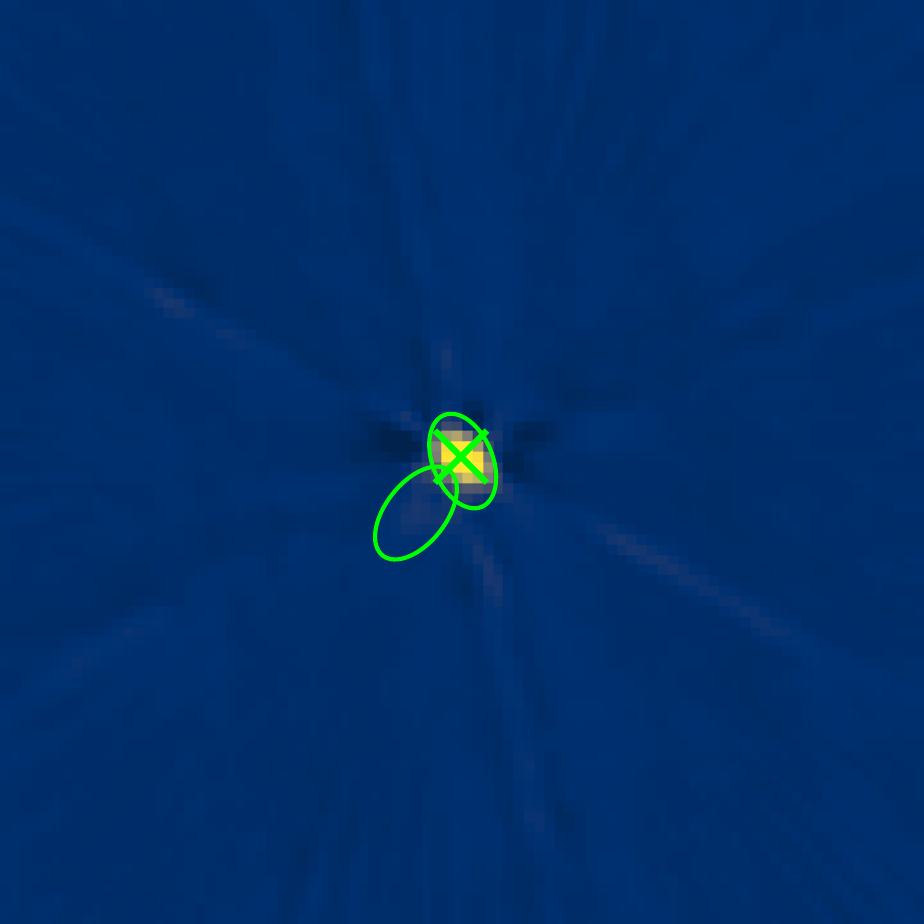}{0.33\textwidth}{(d) J061926.89-302015.2}
        \fig{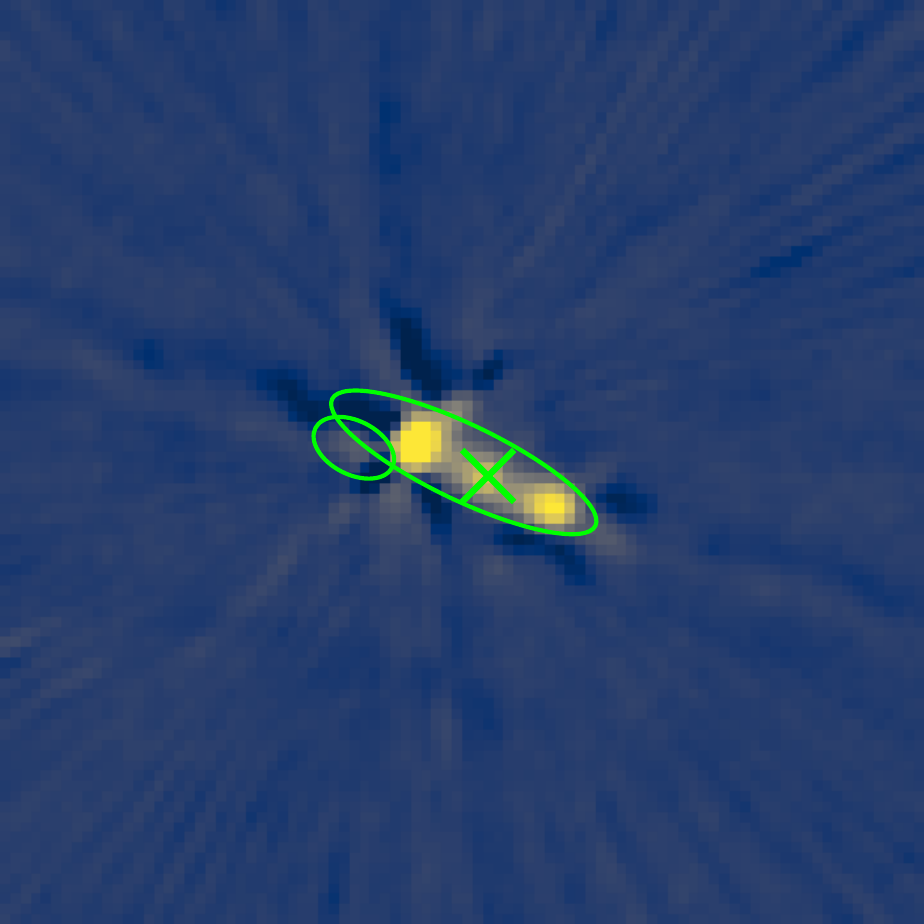}{0.33\textwidth}{(e) J100742.59+590810.6}
        \fig{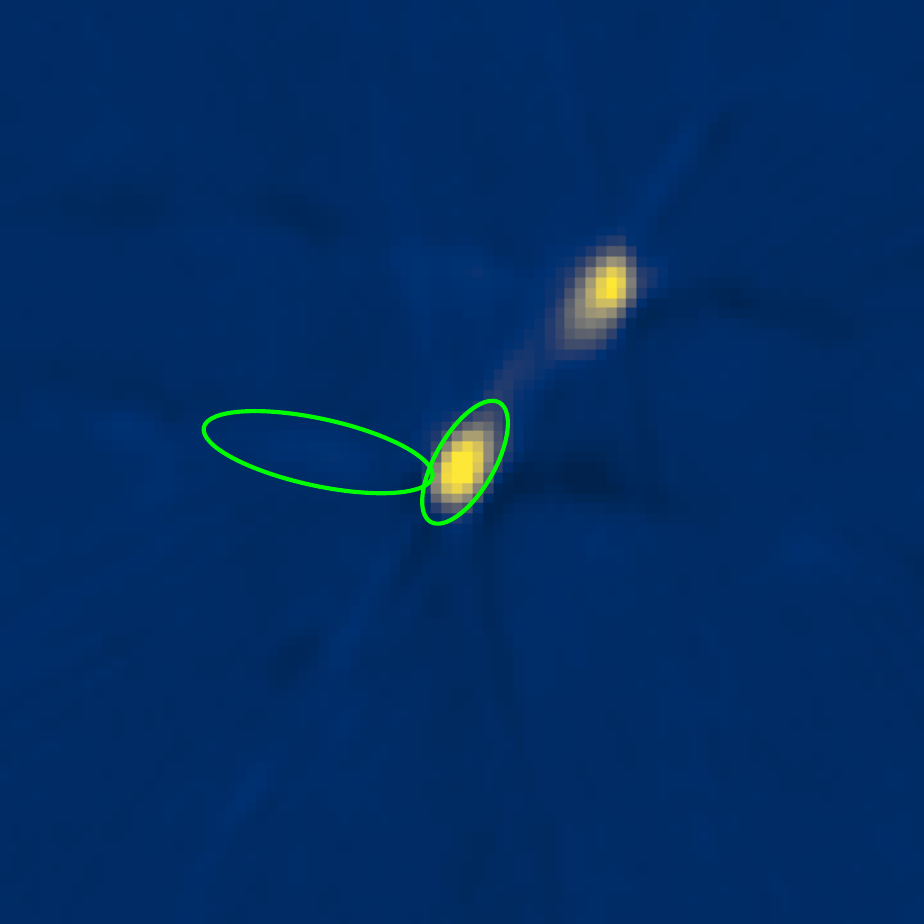}{0.33\textwidth}{(f) J210408.78+763301.1}
        }
    \caption{1.5'x1.5' VLASS cutouts of doubles identified by DRAGNhunter. The ellipses denote components as identified by DRAGNhunter, where the green ellipses denote the lobe or jet hot spot components, and the green X denotes the AllWISE host as identified in \citet{gordon_quick_2023}, if one was found. \\
    (Top line) Examples of what typical 2-artifact doubles look like. Note the similarity between 3-artifact triples in Fig. \ref{fig:triples_a_examples} and 2-artifact doubles: both usually arise in noisy regions around a bright source. \\ 
    (Bottom Line) Examples of what a typical 1-artifact doubles look like. Image (d) represents the majority of 1-artifact sources, however the range of 1-artifact morphologies is larger than that of the triple source set. Images (e) and (f) show how 1-artifact doubles arise also in extended sources with poor component fitting and bright lobes.}
    \label{fig:doubles_a_examples}
\end{figure*}

First, consider a typical 2-artifact triple: a bright, unresolved point source with 2 artifacts around it, e.g. Fig. \ref{fig:triples_a_examples}(d-f). A 1-artifact double, where one component is also a bright, unresolved point source and the second is an artifact, is identical in structure to the 2-artifact triple, i.e. an artifact around a point source. Typical 1-artifact doubles are shown in Fig. \ref{fig:doubles_a_examples}(d-f), which show the near-identical structure of these types of doubles and triples. Now, consider 3-artifact triple sources. These predominantly appear in noisy regions around a single, unresolved point source, e.g. Fig. \ref{fig:triples_a_examples}(a-c). Two-artifact double sources share the same structure, predominantly artifacts around a single bright source, e.g. Fig. \ref{fig:doubles_a_examples}(a-c). Hence, 2-artifact and 3-artifact triple sources are analogous to 1-artifact and 2-artifact double sources.

However, the separation between the clusters is only loose using LAS and flux S/N. Due to the size of the doubles catalog (15,888 sources), visual inspection is impractical with a small team. Thus, we invoke the use of a random forest model, which can identify and learn the much less obvious correlations between parameters and classes and make this classification with more nuance. Using these more granular correlations alongside the already apparent correlations LAS and flux S/N and the number of artifacts, a random forest model is well-poised to identify the number of artifacts in each source. 

Due to the structural similarities between the LAS and flux S/N plots for both subsets in Fig. \ref{fig:las_flux_sn} it is likely that the primary artifact-distinguishing parameters are the same in both datasets. The analogy between double and triple artifact classes further implies that an accurate random forest model trained on triples is likely to be accurate classifying doubles.
%
%==================================================================================%
%==================================================================================%
%
%
%
%
%==================================================================================%
%==================================================================================%
%
\section{Random Forest Methodology} \label{sec:methods}
%==================================================================================%
\subsection{Data and Model Preparation} \label{sec:model_prep}
We used the \verb|RandomForestClassifier| provided in the open source machine learning Python package scikit-learn \citep{pedregosa_scikit-learn_2011} to implement our random forest model. This implementation differs from the implementation in \citet{breiman_random_2001}, where each tree votes for a single class, by taking the average probability of each class taken across each tree. To prepare the data for the classifier, we took Table 6 provided in \citet{gordon_quick_2023} and appended calculations of the LAS signal-to-noise (calculated as LAS/e\_LAS), flux signal-to-noise (calculated as Flux/e\_Flux), and the prominence of the brightest component. These error values are derived from the outputs of PyBDSF, the algorithm used to identify the radio sources in \citet{gordon_quick_2021}, which provides directly the e\_Flux, positional errors, and size errors of each component.

We removed parameters that only triples have from the training set, core prominence and core location data, to ensure that the random forest model trained on the triples dataset can classify sources in the doubles dataset. We also removed the quality flag (SourceFlag, i.e., the `Q' flag) to force the model to focus on the raw data, as the quality flag was assigned based on a synthesis of signal-to-noise and component flux ratio thresholds. Ideally, we want the model to learn a better version of these simple quality thresholds. In order to reduce the number of redundant parameters in the model, we kept only the RA and DEC of the entire source and removed the core position, median position, and flux-weighted position of each source. A detailed description of what each of the parameters are and how they were found is available in \citet{gordon_quick_2023}.

We scaled the training parameters using the \verb|StandardScaler| function available in scikit-learn to standardize the dataset by scaling each feature to have a mean of 0 and a standard deviation of 1. This is common practice in preparing machine learning datasets, and helps prevent outliers and features with different ranges from skewing the training set. Using the triples visual classifications as ground truth, we performed a cross-validating grid search through a range of values for the maximum depth of each tree and number of trees, and we found that a maximum tree depth of 16 and 400 trees provided the best model performance. Therefore, we used these hyperparameter values for all training runs of the random forest classifier. As seen in Table \ref{tab:triples_class_fraction}, 0-artifact sources dominate the population of the triples set and cause a significant imbalance between the 0-artifact and artifact-containing classes. This causes imbalance in the class distribution of our randomly-selected training set. We tried to mitigate the class imbalance of the training dataset by using the `balanced' argument available in the scikit-learn implementation of the random forest classifier. This argument applies weights to each class which are inversely proportional to each class's frequency in the training data and are calculated as eq. \ref{eq:training_class_weight}.

\begin{equation}\label{eq:training_class_weight}
    \text{weight} = \frac{n_{\text{samples}}}{n_{\text{classes}}\cdot n_{\text{occurrences in training set}}}
\end{equation}

It is important to consider this unequal class distribution when evaluating the performance of our random forest model. Raw accuracy, i.e., the fraction of correct classifications, will give an inflated impression of the model's performance on imbalanced datasets \citep{brodersen_balanced_2010}. Thus we use the the F1 score, the harmonic mean of the precision and recall of a model's classifications, which provides a more nuanced account of model performance \citep{chinchor_muc-4_1992}.
 
Precision measures the fraction of predicted positives that are true positives, members of the positive class the model predicts correctly as being a part of the positive class. Precision can be interpreted as the \textit{purity} of the predicted sample of positives, quantifying the level of contamination by false positives, non-members of the positive class that the model incorrectly predicts as being a part of the positive class. Recall measures the fraction of true positives that are correctly identified by the classifier. This quantity can be understood as the \textit{completeness} of the sample of predicted positives, indicating how effectively the model recovers members of the positive class. By combining purity (precision) and completeness (recall), the F1 score captures the compromise between minimizing contamination and maximizing recovery, making it well suited for evaluating performance on imbalanced datasets, and is defined in eq. \ref{eq:F1_score}.

\begin{equation} \label{eq:F1_score}
    \text{F1 Score} = 2\cdot\frac{\text{precision}\cdot\text{recall}}{\text{precision}+\text{recall}} = \frac{2\text{TP}}{2\text{TP} + \text{FP} + \text{FN}}
\end{equation}

In eq. \ref{eq:F1_score} TP refers to true positives, FP refers to false positives, and FN refers to false negatives, where the model fails to classify actual member objects as a part of the class. 

When there are more than just two classification categories, a composite F1 score is found by first calculating the F1 score for each binary classification set that can be constructed from the possible categories, one for each possible label. The F1 score from each of these binary combinations is then averaged to create the final score. Due to the imbalance in the training set, we used the \verb|weighted| preset for the F1 score, which combines each F1 score with a weight like the class weights in eq. \ref{eq:training_class_weight}. Importantly, this score will usually be lower than the actual fraction of correct classifications of a model, however this score is a better metric for comparing performance across different models. Hence, we use the weighted F1 score as the primary measure of model performance and accuracy when evaluating our models.

%==================================================================================%
\subsection{Determining Final Artifact Classification Classes} \label{sec:finding_best_classes}
A perfect random forest model should provide results that match the visual classifications, reliably classsifying the number of artifacts in each source. However, the clustering in Fig. \ref{fig:las_flux_sn}(a) suggests that this may prove difficult, especially for 1-artifact sources (see section \ref{sec:1a}). Therefore, we opted to examine different artifact classifications to determine which would provide the best balance between detail and accuracy. We used the triples visual inspection results summarized in Table \ref{tab:triples_class_fraction} as the ground truth for our random forest model's classification. We randomly selected 20\% of the triples to create a verification set to assess the performance of each model, and the remaining 80\% became the training data set.

\begin{figure*}[htb!]
    \gridline{\fig{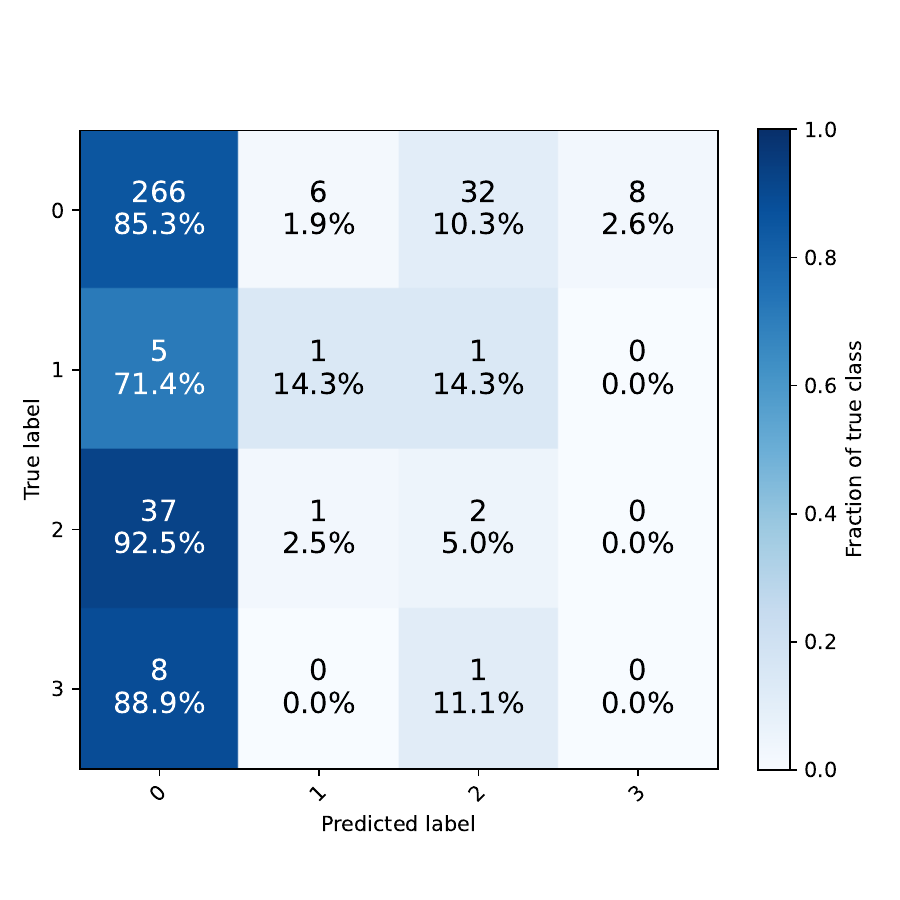}{0.45\textwidth}{(a) Run 1: Control}
            \fig{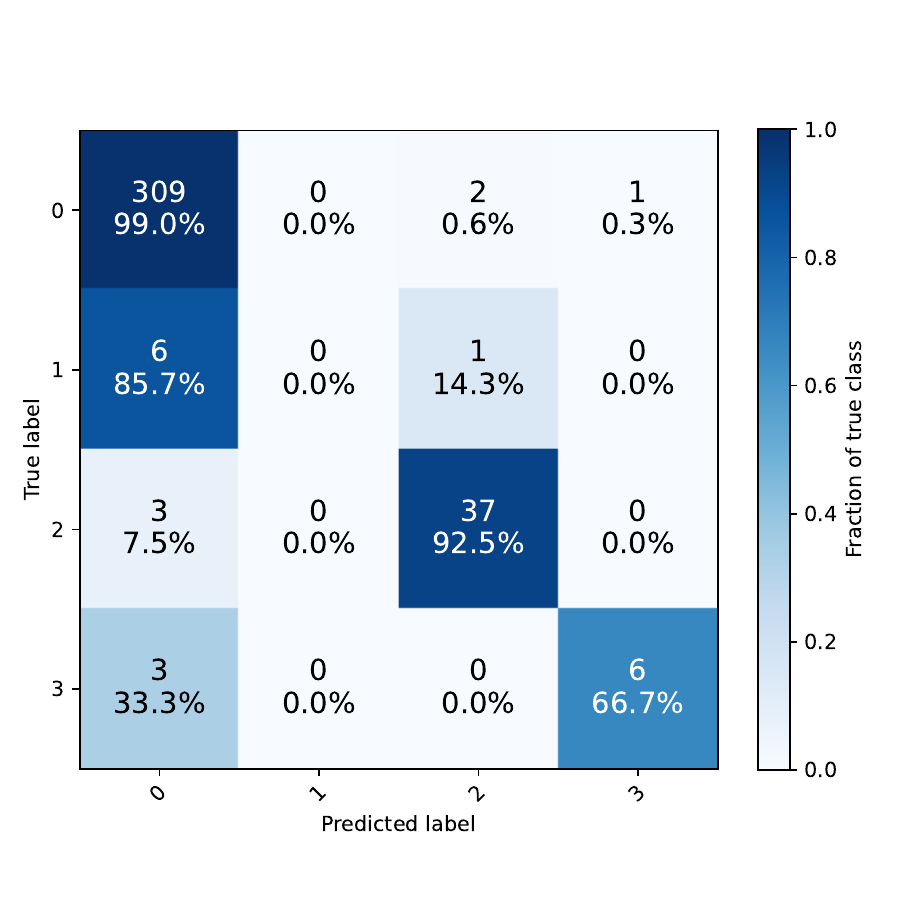}{0.45\textwidth}{(b) Run 2: 0, 1, 2, or 3 artifacts}}
    \gridline{\fig{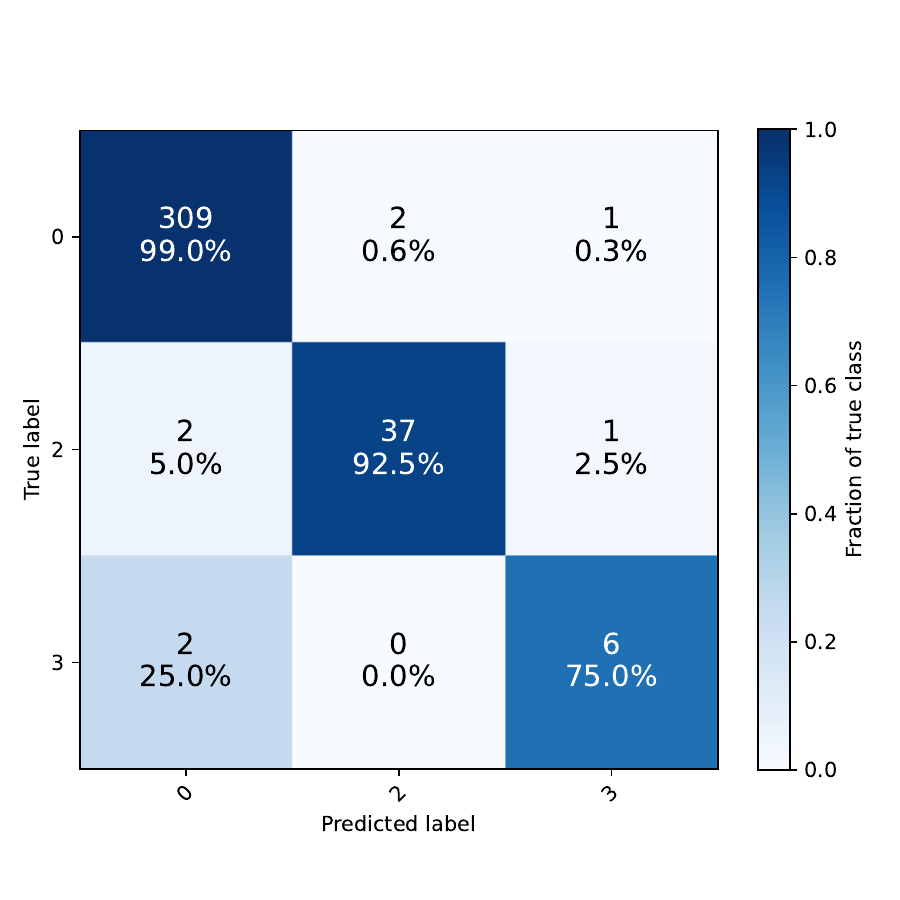}{0.45\textwidth}{(c) Run 3: 0, 2, or 3 artifacts}
            \fig{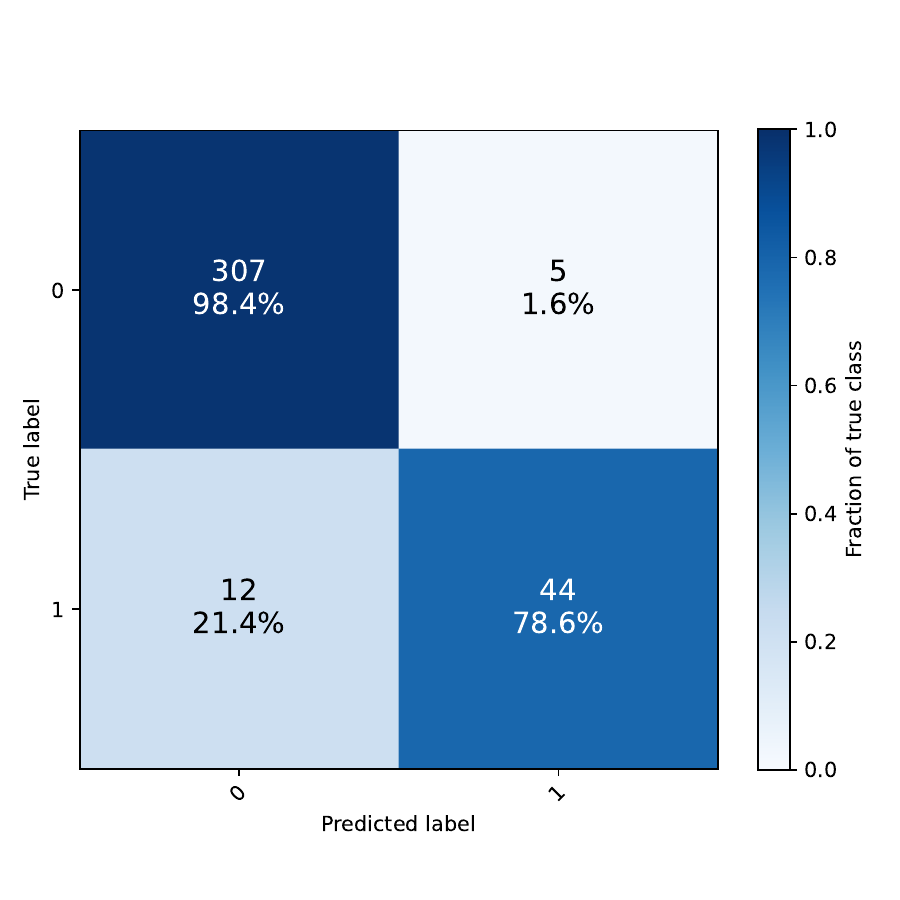}{0.45\textwidth}{(d) Run 4: 0 or any artifacts}}
\caption{Confusion matrices of the classification results on the verification set (20\% of sample) for each of the triples classification runs. The percentages and the color of each square are determined by the fraction of the total population of the row, i.e. the fraction of the true class, that is present in each square.}
\label{fig:artifact_class_test_conf}
\end{figure*}

First, we obtained a control by providing the random forest classifier only the RA of the source and the RA of the core; we call this Run 1. Ideally, sky position should have no significant effect on whether a source has artifacts.\footnote{There are some image quality variations through the VLASS \textit{Quick Look} sky footprint, particularly sources near the lowest declinations observable by the VLA \citep{gordon_quick_2021}, hence, we avoid using the declination in control runs.} In Run 2 we trained the model to identify 0, 1, 2, and 3 artifact triple sources. In Run 3 we reduced the classes to just 0, 2, and 3 artifacts by eliminating all 1-artifacts sources from the training and verification sets. Finally, in Run 4, we treated all classes with 1 or more artifact as `having artifacts' and 0 artifact sources as `not having artifacts' and had the model classify accordingly. As this is only a preliminary test comparing relative accuracy, we only trained a single model for each run. The same verification set was used for all runs except for Run 3, where 1-artifact sources were removed. We show the results of these models in confusion matrices in Fig. \ref{fig:artifact_class_test_conf} which show exactly how well each run classified each artifact class. The runs, the classes they used, and the model's F1 score is summarized in Table \ref{tab:run_results}.

\begin{figure}[htb!]
    \epsscale{.9}
    \plotone{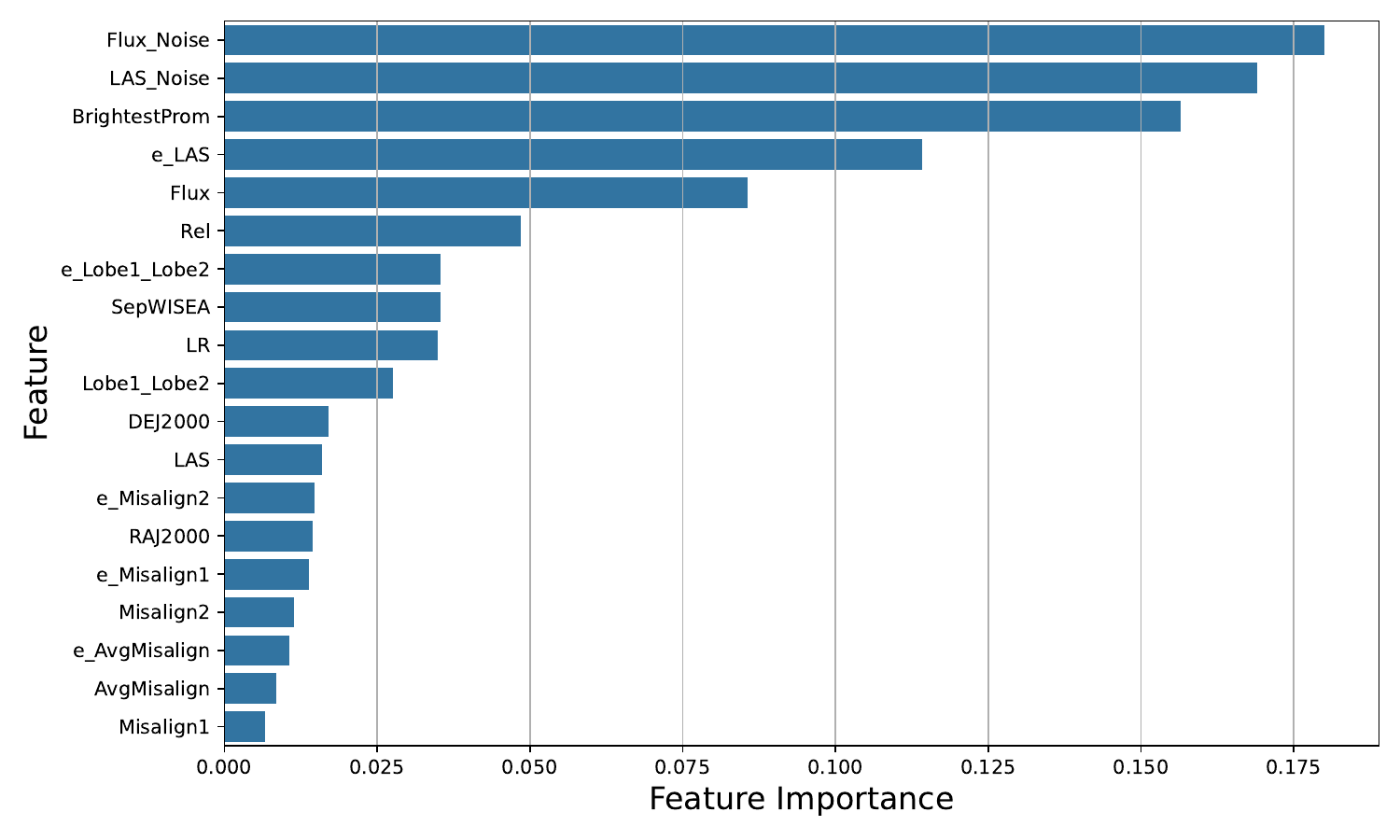}
    \caption{Plot of the importance of each parameter used in the model from Run 3.}
    \label{fig:run_3_importances}
\end{figure}

From Fig. \ref{fig:artifact_class_test_conf}(a) we see that the control run, as expected, did not learn any distinguishing features of the artifact classes and simply classified most objects as 0-artifact because this is the most populous class. Run 2 (Fig. \ref{fig:artifact_class_test_conf}(b)) shows that the model was unable to reliably distinguish 1-artifact objects, and this is discussed in more detail in section \ref{sec:1a}. Runs 3 and 4 (Fig. \ref{fig:artifact_class_test_conf}(c-d)) show the most promise due to very similar F1 scores and confusion matrices, which show they reliably distinguish members of the classes they were trained to identify. We verify the performance of Run 3, which provides the best compromise between level of detail and accuracy, by checking which features were most useful for distinguishing classes correctly, i.e. the importance of each feature, in Fig. \ref{fig:run_3_importances}.

The model found the LAS S/N and flux S/N as well as the prominence of the brightest component to be the most important features for distinguishing classes correctly, consistent with our findings in section \ref{sec:motivation}. The model learned the real separating features of the dataset and combined them with the other catalog features to produce classifications with accuracy comparable to manual visual classification. We choose the artifact classes in Run 3, 0-, 2-, and 3-artifact, as the artifact classes of our random forest model in subsequent sections because these classes provide the best compromise between classification level of detail and accuracy.

\begin{deluxetable}{lcc}[htb!]
    \tablecaption{Description of the runs used to determine which set of artifact classes should be used in the triples-trained model.}
    \label{tab:run_results}
    \tablehead{
        \colhead{Run No.} & \colhead{Class Description} & \colhead{Weighted F1 Score (\%)}
    }
    \startdata
    1 & 0-3 artifacts; control & 72.7 \\
    2 & 0-3 artifacts; relevant catalog parameters & 94.7 \\
    3 & 0, 2, or 3 artifacts & 97.8 \\
    4 & 0, or 1 or more artifacts & 95.3
    \enddata
\end{deluxetable}

\subsubsection{1-Artifact Triple Sources} \label{sec:1a}

As seen in Fig. \ref{fig:artifact_class_test_conf}(b), the random forest model was completely unable to classify 1-artifact triples. There are only 37 total 1-artifact triple sources and they comprise such a small fraction of the triples set that both the training and verification sets are starved of them. In ideal cases, random forest models do not need large populations to learn the properties of a class and provide reliable classifications, such as with the subset of 3-artifact triples which make up only 2.3\% of the dataset. As shown later in Section \ref{sec:training_sel}, random forest models rely on parameters that distinguish between classes, such as LAS S/N, flux S/N, and the prominence of the brightest component in the VLASS DRAGNs catalog. The lack of reliable 1-artifact classifications would be mitigated if 1-artifact sources had a distribution in parameter space that makes them distinct from the other artifact classes, however they masquerade as both artifact-free triples and 2-artifact triples as seen in Fig. \ref{fig:las_flux_sn}. We found that no parameter provided in the VLASS DRAGNs catalog, nor any additional parameter we derived from this catalog reliably distinguishes 1-artifact triples from the other artifact classes. 

1-artifact triples triples arise mainly from extended sources with one particularly bright lobe, which tends to cause artifacts. The prevalence of these bright lobes explains how some 1-artifact triples occupy the same area in LAS S/N and flux S/N space as 2-artifact sources, because their morphology is similar. Some 1-artifact sources subdivide what should be single components into multiple detections, which will be similar to the other components and this helps to explain why the rest of the 1-artifact sources align with the majority of the rest of the triples population. These morphological similarities to other classes and lack of distinguishing parameters are what prevent us from reliably classifying 1-artifact triples. This is not an issue for random forest classification of the doubles set, as the mapping of triples artifact classes to doubles discussed in Section \ref{subsec:motivation_analogy} relies only on 0-, 2-, and 3-artifact triples, which we have shown already that the model can reliably classify.

%
%
%
%==================================================================================%
\subsection{Training Set Selection} \label{sec:training_sel}
While random forest models are resistant to overfitting, they converge to an accuracy limit as the size of the training set increases. \citep{ho_random_1998,breiman_random_2001}. It is important to understand how large the training set for a dataset should be, and the minimum size needed to achieve desired performance. This can be difficult to do initially, and the selection method for constructing the training set can affect significantly how large or small the set should be. The traditional method is random selection, and in this section we compare this method's performance to our own selection method which leverages the LAS S/N and flux S/N clustering identified in Fig. \ref{fig:las_flux_sn}. 

\begin{figure}[htb!] 
    \epsscale{1}
    \plotone{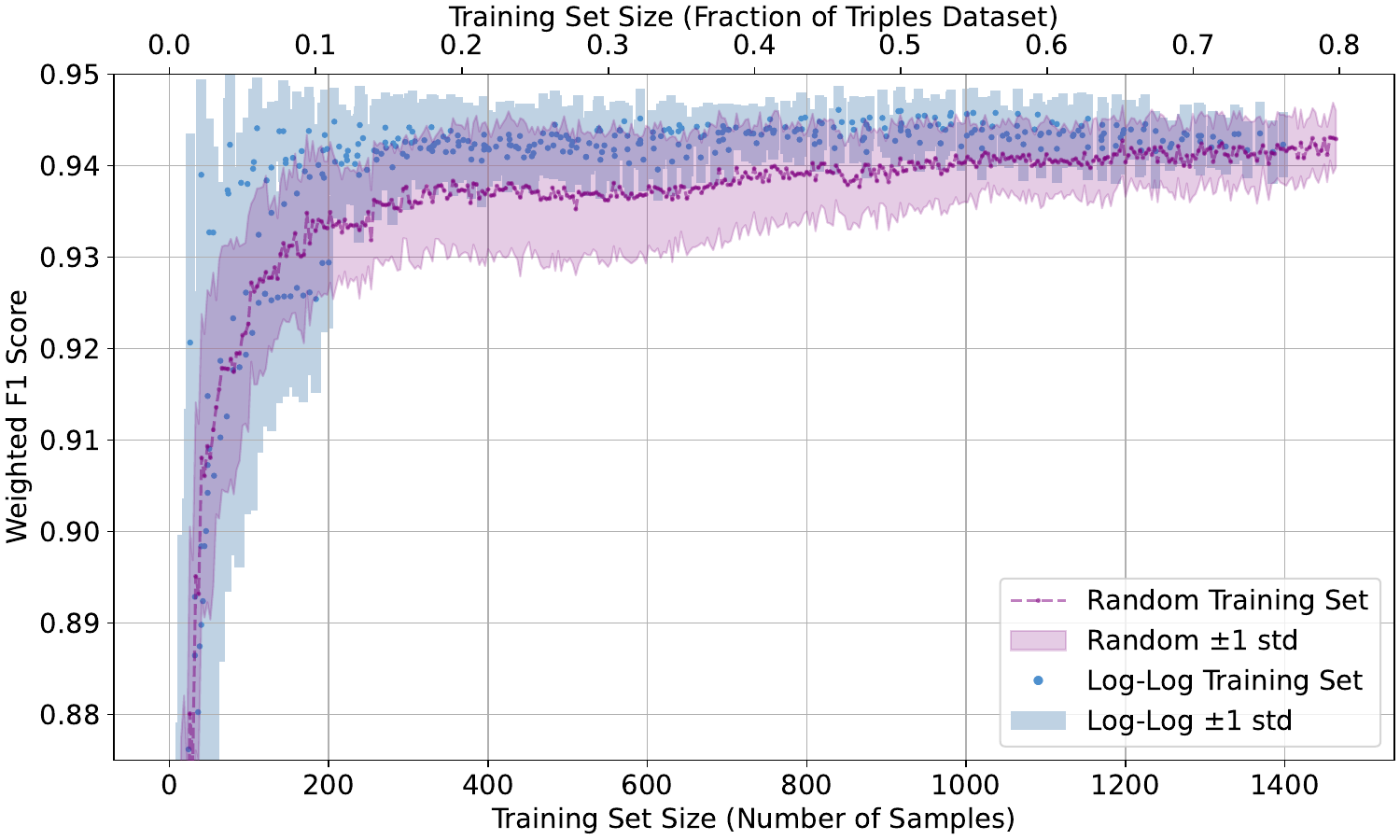}
    \caption{Plot of the mean weighted F1 score across 25 randomly seeded runs per point for both random and log-log parameter space training set selection methods. The log-log method using LAS and Flux S/N converges towards the maximum accuracy sooner than with random selection. We performed additional runs with log-log selection varying from 1-15 bins per axis and 30-50 samples per bin to reach approximately the same maximum sample size as with the random selection method. Especially for smaller training set sizes, runs with fewer bins but more samples per bin performed more poorly than runs with more bins but fewer samples per bin.}
    \label{fig:random_vs_loglog}
\end{figure}

We took our triples dataset and randomly selected 20\% of sources to construct a representative verification set. This set was kept consistent through all subsequent tests of training size and selection methods. First, we trained random forest models on training sets that were selected randomly with varying size, and we evaluated performance using the average weighted F1 score across 25 different randomly-seeded random forest runs at each training set size. We plot the results of this in Fig. \ref{fig:random_vs_loglog} in purple. The model's accuracy decreases in standard deviation and increases in average accuracy until it reaches the maximum possible training set size where the training set is comprised of all of the remaining triples not in the verification set.

One reason for the continual, gradual increase in performance of the model as the size of the random training set increased is the class imbalance of the dataset. Models with lower training set size select only a few sources that contain artifacts, missing artifact-containing sources which possess greater learning value to the model because they help the model to make meaningful distinctions between artifact classes. The most straightforward method to combat this imbalance is to construct the training set strategically such that it is less imbalanced. Though this means that the training set is not representative, one can take first a representative sample out of the entire set and set it aside as a verification set. From our exploration of data clustering in Section \ref{subsec:prelim_clustering}, and particularly Fig. \ref{fig:las_flux_sn}, we know that there is clustering in LAS S/N and flux S/N parameter space, and we can use this as a basis for selecting sources which are more likely to contain artifacts. Doing so will allow us to selectively sample the parameter space to improve representation of minority class members and reduce the proportion of majority class members, i.e., increase the number of artifact-containing sources in the training set. This data-motivated up-sampling method allows us to construct training sets which achieve the same performance as larger training sets constructed by random selection. 

\begin{figure*}[htb!]
    \gridline{
    \fig{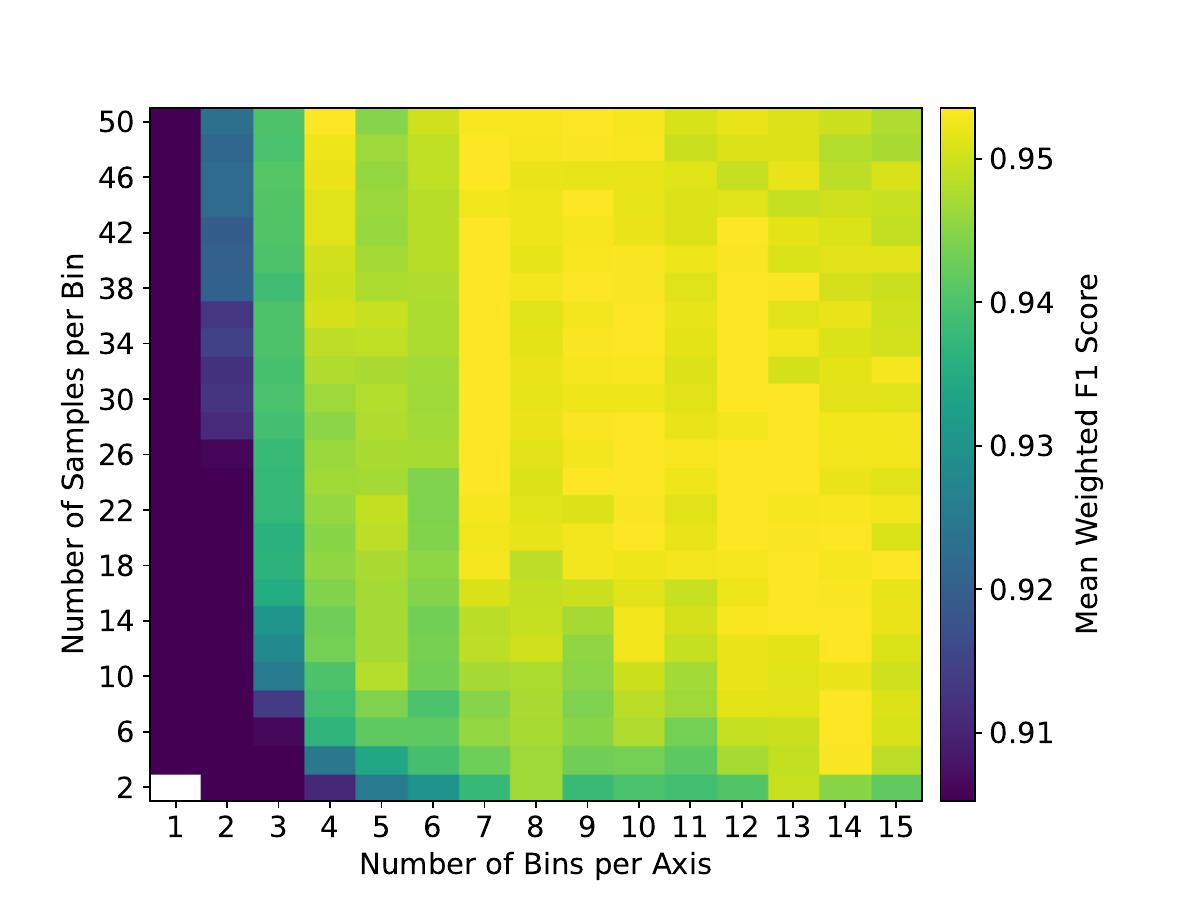}{.5\textwidth}{(a)}
    \fig{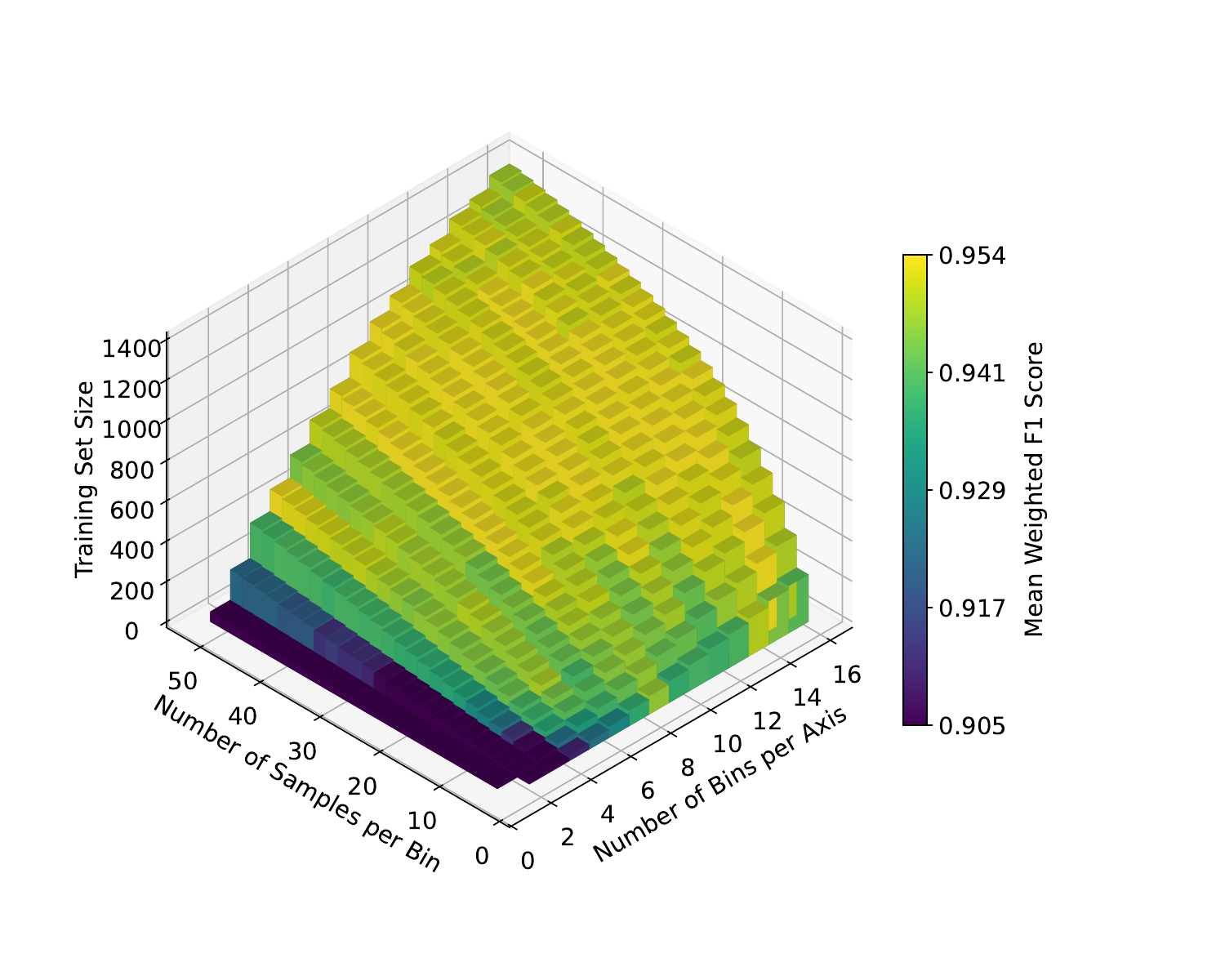}{.5\textwidth}{(b)}
    }
    \caption{Plots comparing the performance of models trained on sets selected by log-log LAS S/N and flux S/N parameter grid space selection across different combinations of bins per axis and maximum number of samples per bin. Because we define a uniform grid of bins, and our sources are not uniformly distributed across LAS S/N and flux S/N parameter space (as seen in Fig. \ref{fig:las_flux_sn}), some bins contain less than the maximum number of sources per bin, and only sample as many as are contained within the bin. Each combination of selection parameters was tested across 25 differently-seeded random forest models. (a) Heatmap of accuracy across the selection parameters. (b) 3D bar plot extending (a) with the height of columns corresponding to the actual size of each training set. In both panels, the F1 scores were clipped to the 10-90th percentile range and linearly mapped to the colormap; values outside this range are saturated, particularly for sets with very few bins per axis where the actual F1 score is between 0.8-0.9.}
    \label{fig:loglog_tuning}
\end{figure*}

We define this method, log-log parameter grid space selection, by constructing a grid of equal width bins in logarithmic parameter space and randomly selecting up to a certain maximum number of samples from each bin. This method works because we are sampling across the entire parameter space and thus still represent the full range of the parameter data while improving the training set class distribution. Because there is clustering in LAS S/N and flux S/N at the extrema of the parameter space, we can apply this selection method to prioritize selecting 2- and 3-artifact sources in our training set. In order to determine the number of bins and the number of samples per bin, we iterated through an array of number of bins and number of samples and found the average weighted F1 score across each set of 10 iterations at each combination of values. The results are shown in the heatmap in Fig. \ref{fig:loglog_tuning}(a) and are extended to also show the actual size of the training dataset in Fig. \ref{fig:loglog_tuning}(b). As with our testing with random training set selection, we trained 25 randomly-seeded models for each combination of bins and samples and used the average weighted F1 score to measure performance.

In Fig. \ref{fig:random_vs_loglog} we plot the results of log-log parameter space selection in blue alongside the results of random selection. Comparing these methods we find that the former reaches the maximum possible accuracy of the model with significantly fewer samples. This implies that our log-log parameter space selection method is more effective than random selection because only a comparatively small training set is needed to achieve the same performance. This is possible exactly because of the clustering we identified in Section \ref{sec:motivation}, and if this parameter space clustering exists for spurious detections in other surveys then this sampling method is a promising approach to construct less imbalanced training sets more generally.

%
%==================================================================================%
%==================================================================================%
%
%
%
%
%==================================================================================%
%==================================================================================%
%
\section{Random Forest Classification of Doubles} \label{sec:doubles_class}
\subsection{Classification Models}
We classified the doubles with two different models: 1) the most effective version of the triples random forest classifier, the model from Run 3 in Section \ref{sec:finding_best_classes}, and 2), a model trained on a small subset of doubles created with log-log LAS S/N and flux S/N parameter grid space selection. As discussed in section \ref{subsec:motivation_analogy}, the classes in the triples dataset are analogous to the classes in the doubles dataset, so we can map the classifications of the triples-trained model of 0, 2, and 3 artifacts to 0, 1, and 2 artifacts in doubles with confidence. We compare the triples-trained model's performance to the classifications of a model trained on a set selected by log-log parameter grid space selection to determine if a model trained on a minimal training set of doubles can attain performance similar to the triples-trained model.

In order to evaluate the performance of both models, we randomly selected 600 doubles and classified them by number of artifacts using the criteria discussed in Section \ref{subsec:motiv_classification}. Every source was classified by multiple members of the team, and final classifications were reached by group consensus. This set is our verification set for assessing the performance of models applied to the doubles population, and we take this as a representative sample of the entire doubles population. The artifact class distribution of doubles is provided in Table \ref{tab:doubles_inspected_class}. Because over 90 percent of sources are 0-artifact, compared to approximately 85 in the triples, we see that the class distribution is more biased towards 0-artifact sources in the doubles subset.

\begin{deluxetable}{lccc}[htb!]
    \tablecaption{Class fractions of doubles subsets}
    \label{tab:doubles_inspected_class}
    \tablehead{
        \colhead{Number of Artifacts} & \colhead{0} & \colhead{1} & \colhead{2}
    }
    \startdata
    \cutinhead{Randomly-selected verification set}
    Number & 557 & 34 & 9 \\
    Percent of Total & 92.8 & 5.7 & 1.5 \\
    \cutinhead{Log-log LAS S/N and flux S/N training set}
    Number & 147 & 38 & 19 \\
    Percent of Total & 72.1 & 18.6 & 9.3
    \enddata
\end{deluxetable}

\subsubsection{Triples-Trained Model}
\begin{figure}[htb!]
    \epsscale{.6}
    \plotone{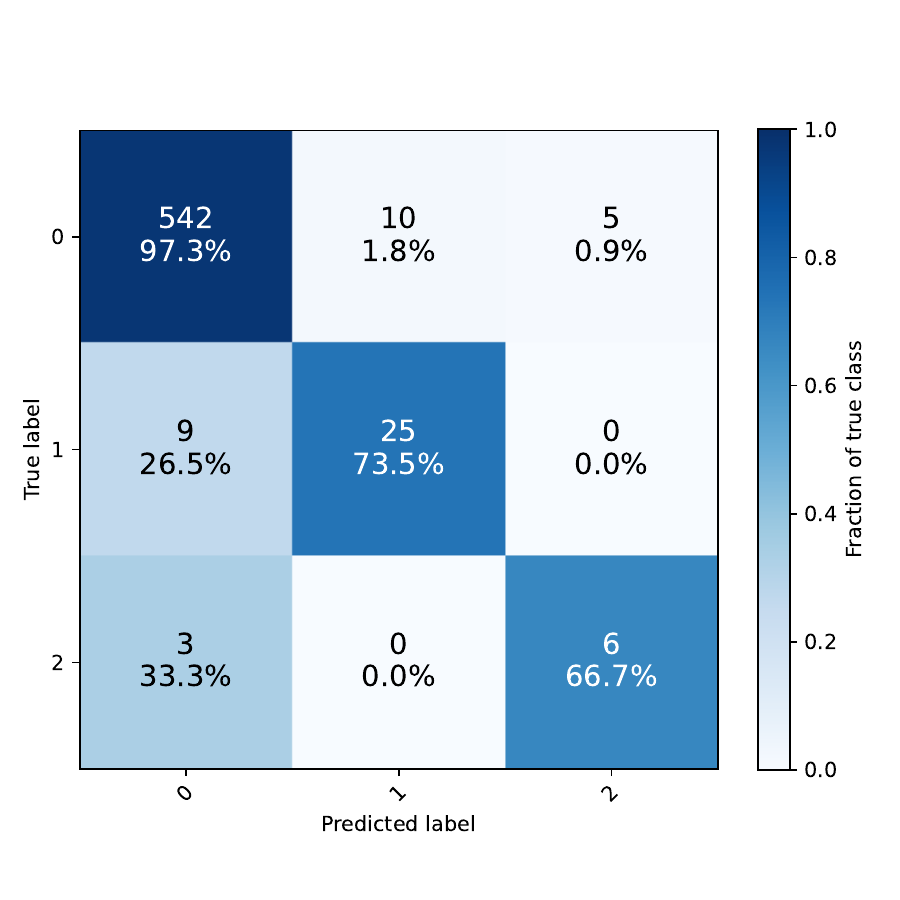}
    \caption{Confusion matrix comparing the predictions of the triples-trained RF model to the results of visual inspection on the doubles verification set.}
    \label{fig:600verif_matrix}
\end{figure}

The triples-trained model was trained on the remaining fraction of sources not in the triples verification set, a random subset of 1,468 triples from the 1,836 total triples. We applied this model to predict the artifact classes of each source in the doubles verfication set, and the confusion matrix of the model's classifications is in Fig. \ref{fig:600verif_matrix}. The model achieved a weighted F1 score of 95.6\%. While this is lower than the weighted F1 score of the same model on the triples verification set, the model is still well-suited for removing contaminated images.

\subsubsection{Log-Log LAS S/N and Flux S/N Selected Model}
We tested the efficacy of using log-log LAS S/N and flux S/N grid space selection to select a minimal training set that can still achieve high classification performance. Our testing of selection parameters, the number of bins per axis and the number of samples per bin, in Section \ref{sec:training_sel} and plotted in Fig. \ref{fig:loglog_tuning} shows that using 6 bins per axis and 8 samples per bin allows the model near-peak performance with a small training set of just over 200 sources. Using log-log LAS S/N and flux S/N grid space selection with these parameters we created a training set of doubles with 204 sources which we visually classified for number of artifacts. As with the triples and the doubles verification sets, multiple people classified each source and came to a consensus about each classification. The resultant class fractions are compared to the class fractions of the doubles verification set in Table \ref{tab:doubles_inspected_class}, verifying that this selection method mitigates part of the class imbalance inherent to the doubles population. 

\begin{figure}[htb!]
    \epsscale{.6}
    \plotone{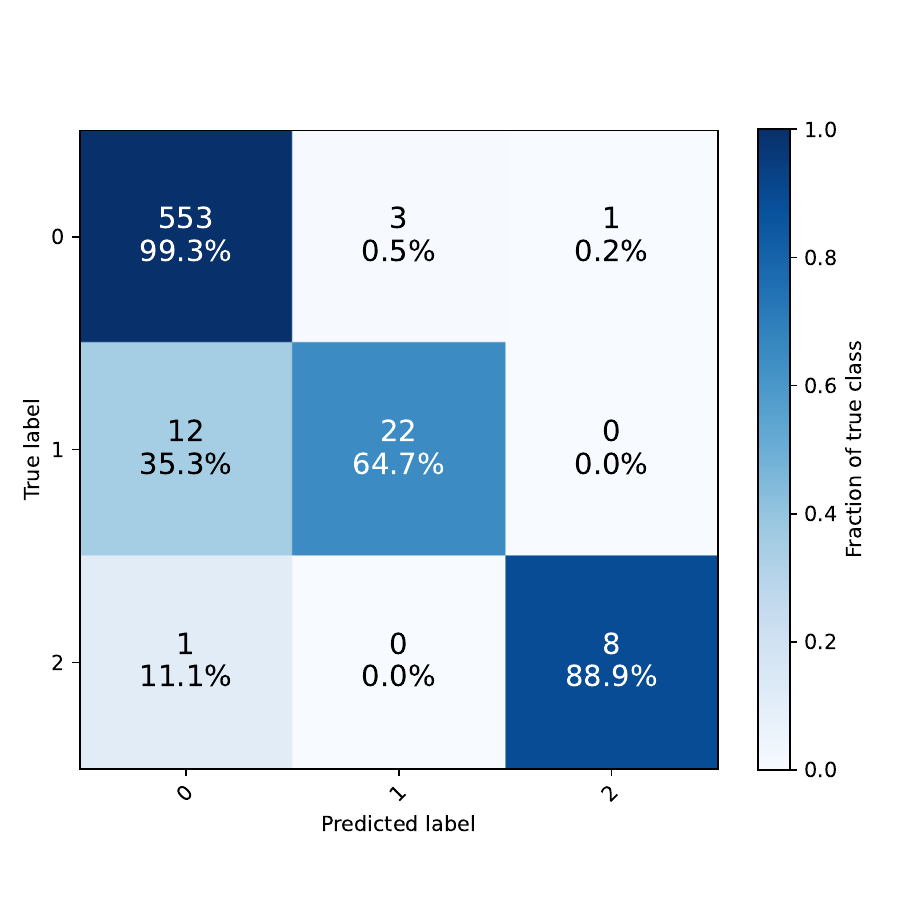}
    \caption{Confusion matrix of the predictions of the model trained on the log-log LAS S/N and flux S/N set of doubles applied to the doubles verification set.}
    \label{fig:loglog_matrix}
\end{figure}

The model achieved a weighted F1 score of 97.0\% when applied to the verification set, and the confusion matrix of the results is shown in Fig. \ref{fig:loglog_matrix}.

\subsection{Doubles Classification Performance and Model Comparison} \label{sec:model_performance}
The weighted F1 score of the log-log model is 1.4\% higher than the triples-trained model, and the log-log model produces fewer false positives, i.e. real 0-artifact sources classified as containing 1 or more artifacts. The triples-trained model confused a higher number of 0-artifact sources as containing artifacts, and this is a consequence of morphology patterns in the doubles population that are not as prevalent in the triples population. In particular, 1-sided 0-artifact sources with bright cores and lobes containing more diffuse emission confused the model and were more frequently classified as 1-artifact sources. More generally, sources with diffuse emission in one or more of the components are more common in the doubles set and were frequently classified as containing artifacts. These sources were a problem for both the triples-trained and log-log models.

This is due in part to how most triples more commonly have well-defined and clearly separated cores and components, a consequence of the source-finding methods of the VLASS catalog from \citet{gordon_quick_2021}. For a core to be identified correctly, there must be clear separation between the lobes and the core, which precludes most scenarios where a component encompasses both the core and a lobe, or the entire extended source. DRAGNs with more continuous emission that is contiguous with their core, or have dim cores, are less likely to have been separated into different sources, i.e. a separate core and two separate jet lobes. These poorly-defined doubles, as well as doubles with a bright lobe and a dim lobe, tend to confuse the models due to how dim lobes and lobes with weak, extended emission can appear similar to artifacts. The triples-trained model erroneously classified 0-artifact doubles with these types of features more frequently than the log-log model. Both models performed best with well resolved doubles and sources that lack components containing more diffuse emission, and these well-behaved sources make up the majority of the doubles. This feature of the doubles set means that the triples-trained model is still able to reach high classification accuracy despite the increased prevalence of sources with dubious morphologies.

\begin{figure}[htb!]
    \epsscale{.8}
    \plottwo{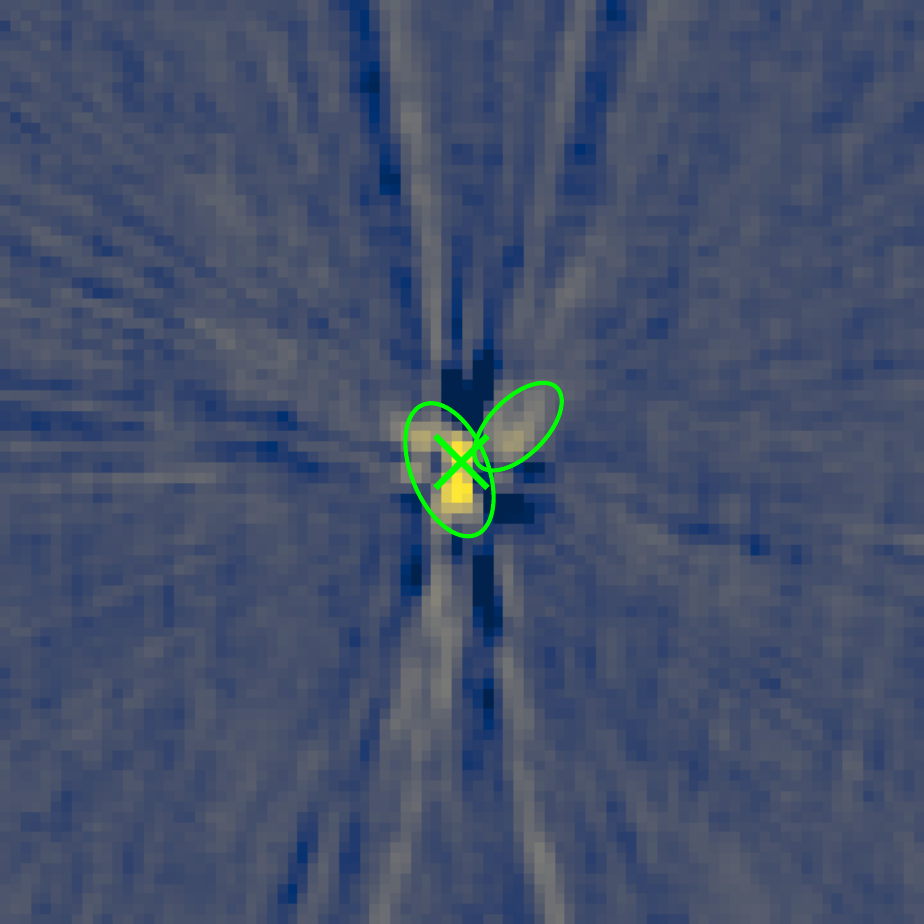}{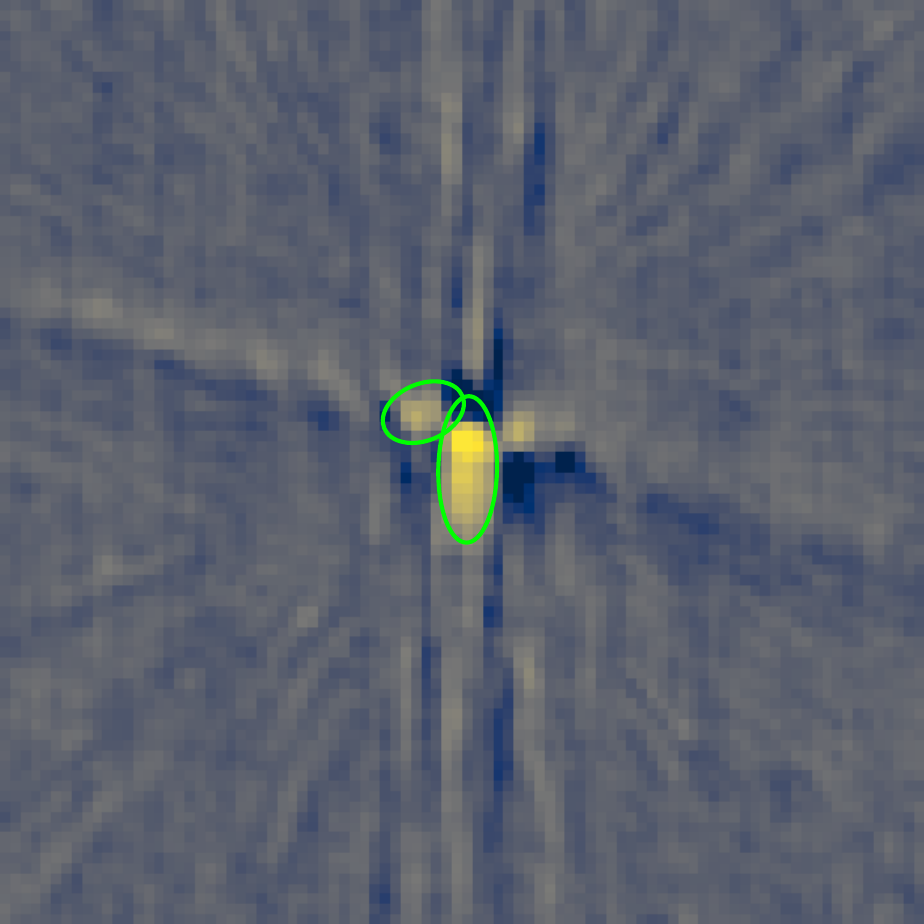}
    \caption{A couple of the sources that confused both the triples-trained and log-log LAS S/N and flux S/N models. The green ellipses denote components as identified by DRAGNhunter, and the green X denotes the AllWISE host as identified in \citet{gordon_quick_2023}, if one was found. Both are unresolved point sources with prominent sidelobes and represent sources that both models misclassified as 0-artifact when they are actually 1-artifact.}
    \label{fig:discrepant_sources}
\end{figure}

The model trained on the log-log LAS S/N and flux S/N set of doubles was more resilient to the 0-artifact classification errors that the triples-trained model was prone to. This is desirable, as the final 0-artifact catalog resulting from this model will be more complete than that of the triples-trained model. However, the log-log model incorrectly classified a greater number of 1-artifact sources as 0-artifact than the triples-trained model, and the resultant 0-artifact set is less pure. These sources contain either prominent sidelobes or bright, high-noise artifacts, the latter of which was an issue for the log-log trained model because the doubles set contains a greater number of 0-artifact sources with dim components and extended emission that look like these bright artifacts. The higher prevalence of these more confusing 0-artifact sources biases the model to be more conservative when classifying sources as 1-artifact. It should be additionally noted that these types of sources were difficult to classify for both models, as all 0-artifact sources the triples-trained model classified as 1-artifact were given the same incorrect classification by the log-log model. This indicates that many of these sources are fundamentally difficult to distinguish with just the parameters given in the DRAGNs catalog. Indeed, these sources were among the most difficult to visually classify. Examples of these sources are provided in Fig. \ref{fig:discrepant_sources}.

More generally, the triples-trained model has a higher false positive rate but a lower false negative rate, i.e., the model tends to classify 0-artifact sources as artifact-containing more frequently, while making fewer incorrect 0-artifact classifications on artifact-containing sources. In contrast the model trained on the log-log selected doubles set has a higher false negative rate but a lower false positive rate. This suggests that these models are brushing against the maximum accuracy limitations of the doubles set, which has a softer morphological distinction between 0-artifact and 1-artifact sources due to doubles with one bright lobe and one faint lobe. 

\subsection{Final Classifications}
Due to the higher F1 weighted score of the log-log model, and that it produces a more complete set of 0-artifact sources, we selected this model to provide the `best' artifact classifications of the doubles set. These classifications are provided by taking the class with the highest predicted probability as predicted by the ensemble of trees in the random forest classifier. These probabilities can be useful in tuning the final classifications of the model, as some sources lie on the border of being classified as one or another artifact class. More pure sets of 0-artifact sources can be obtained by using probability thresholds that lead to less conservative classifications of sources containing artifacts, though this is at the cost of losing 0-artifact sources. 

The triples-trained model provides a more pure set of 0-artifact sources at the cost of completeness, and similar accuracy as the log-log model. As such in our final catalog we provide the class probabilities for each source from both models in addition to the `best' class as provided by the class with the highest probability from the log-log model.

We estimate the error in the best classifications using bootstrap resampling with replacement, a reliable method for calculating the statistics of a population when only a small sample is known \citep{efron_bootstrap_1979, diciccio_bootstrap_1996}. We find, with bootstrapping, that the log-log model using the `best' classifications achieves an F1 score of $97.01\%^{+1.12\%}_{-1.32\%}$, where uncertainties correspond to the 5th and 95th percentiles of the bootstrap distribution. This indicates high performance of our log-log random forest classifier and that our final artifact classification catalog is reliable.  

\subsubsection{Comparison to DRAGNhunter Source Quality Flag}
We compare the existing, more straightforward source-filtering approach used in the VLASS DRAGNs catalog with our random forest method by evaluating the purity and completeness of the set of non-spurious sources identified by each model in the doubles verification set. A perfect source-filtering method would yield of set of non-spurious sources with 100\% purity and 100\% completeness, i.e., be comprised entirely of non-spurious sources and contain all non-spurious sources present in the original catalog. \citet{gordon_quick_2023} defines a source quality flag, `Q', (SourceFlag in column 27 of Table 6) which is intended to filter out the majority of spurious DRAGN identifications based on quantitative thresholds on the LAS S/N and the ratio of flux between the DRAGN lobes. The Q flag provides a higher purity set of the DRAGNs catalog at the expense of eliminating a small number of real sources. This method shares similarity to our random forest model, which, as shown in Fig. \ref{fig:run_3_importances} also relies on the LAS S/N as one of the most important parameters for identifying the artifact class of a source. Using the representative sample of 600 doubles we visually classified, the Q flag filtering approach achieves a completeness of 94.4\% and a purity of 98.8\%.

\begin{figure*}[htb!]
    \epsscale{1}
    \gridline{
    \fig{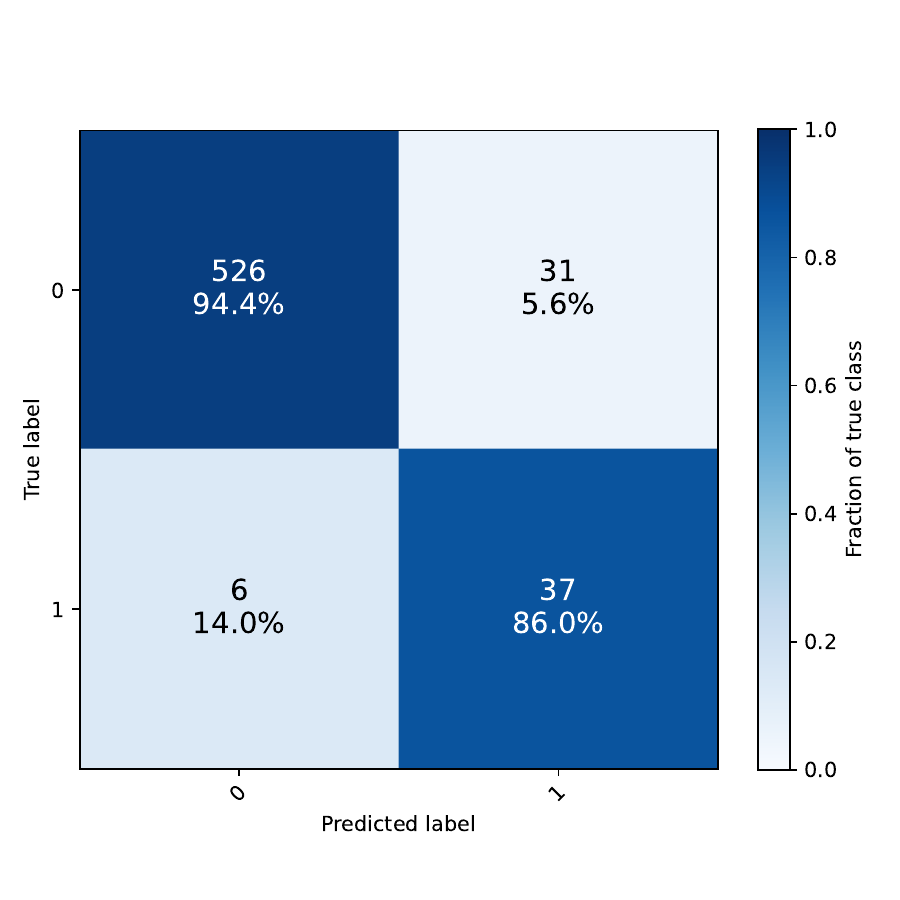}{0.5\textwidth}{(a) DRAGNhunter Q flag}
    \fig{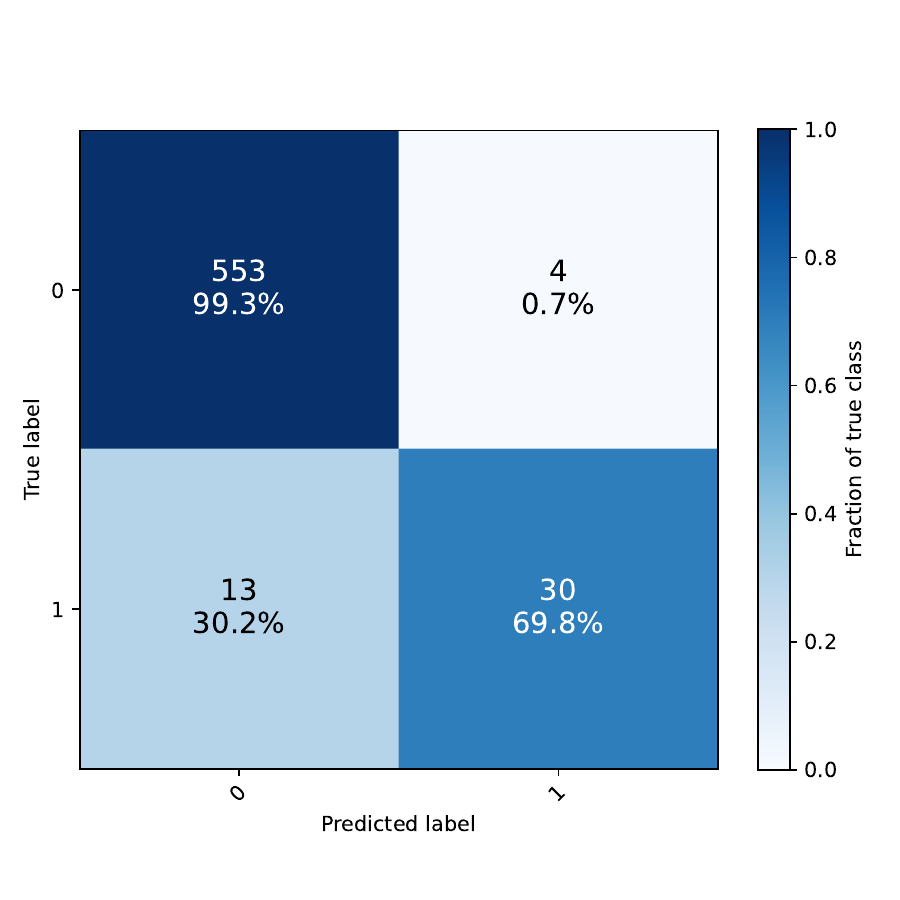}{0.5\textwidth}{(b) Log-log random forest model}
    }
    \caption{Confusion matrices comparing the efficacy of the DRAGNhunter Q flag and our log-log random forest model in isolating spurious DRAGNs, where non-spurious sources are labeled with a 0 and spurious sources are labeled with a 1. While the random forest model misses some artifact-containing sources filtered out by the Q flag, it retains a larger number of artifact-free sources.}
    \label{fig:dragnhunter_conf_comp}
\end{figure*}

Using the set of sources identified by the log-log model as having a `best' artifact class of 0, our random forest method achieves a completeness of 99.3\% and a purity of 97.7\%. We can visually compare the Q flag and random forest approaches with confusion matrices. Because the Q flag is binary and set to 1 if a source is potentially spurious, we map all artifact classifications in the verification set and log-log classifications of 1 or 2 artifacts to 1. These confusion matrices are shown in Fig. \ref{fig:dragnhunter_conf_comp}, and they exemplify why the Q flag results in a higher purity set, but the set classified by our random forest is much more complete. Though the Q flag seems better at identifying spurious sources, it is far less effective in correctly identifying non-spurious sources. 

\begin{figure*}[htb!]
    \epsscale{1}
    \gridline{
    \fig{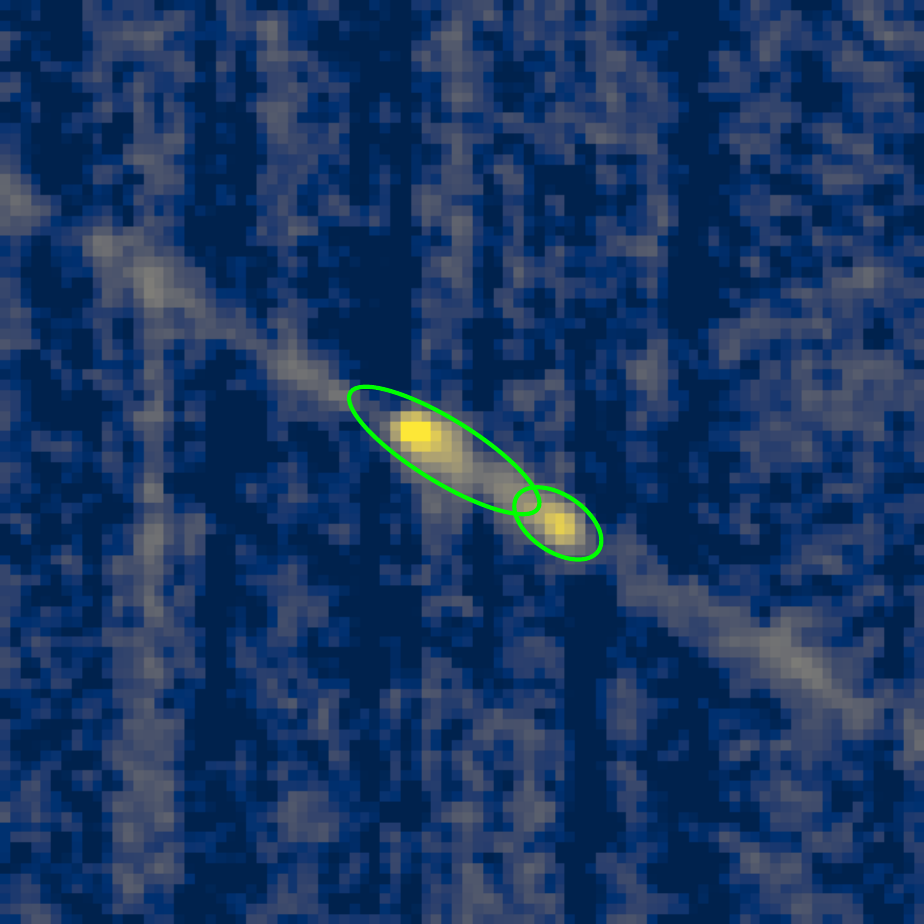}{0.33\textwidth}{(a) J013700.52+481905.4}
    \fig{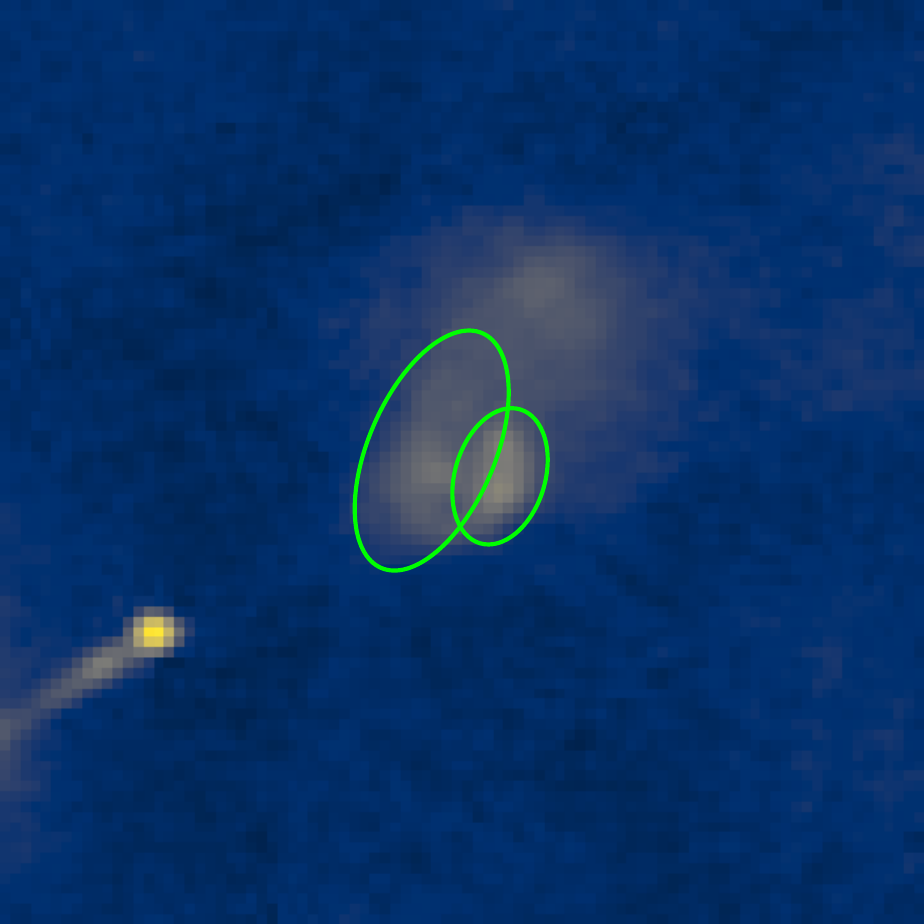}{0.33\textwidth}{(b) J223645.98+385333.6}
    \fig{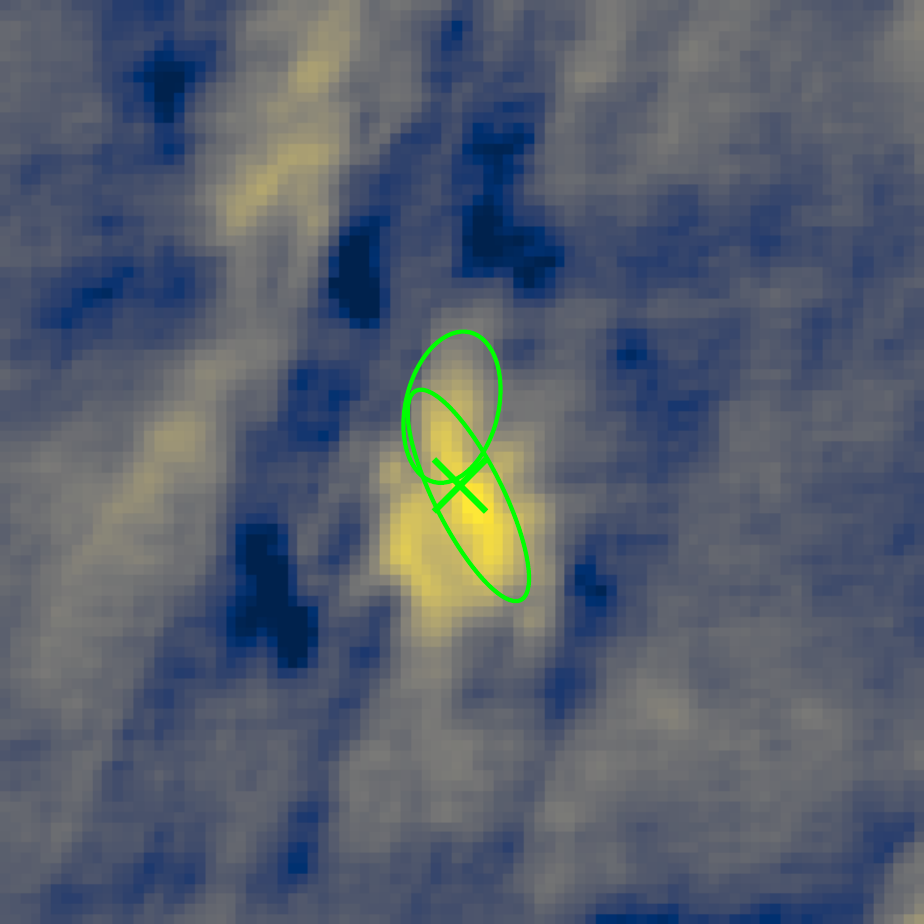}{0.33\textwidth}{(c) J180900.31-200505.5}
    }
    \gridline{
    \fig{./J100742.59+590810.6.pdf}{0.33\textwidth}{(d) J100742.59+590810.6}
    \fig{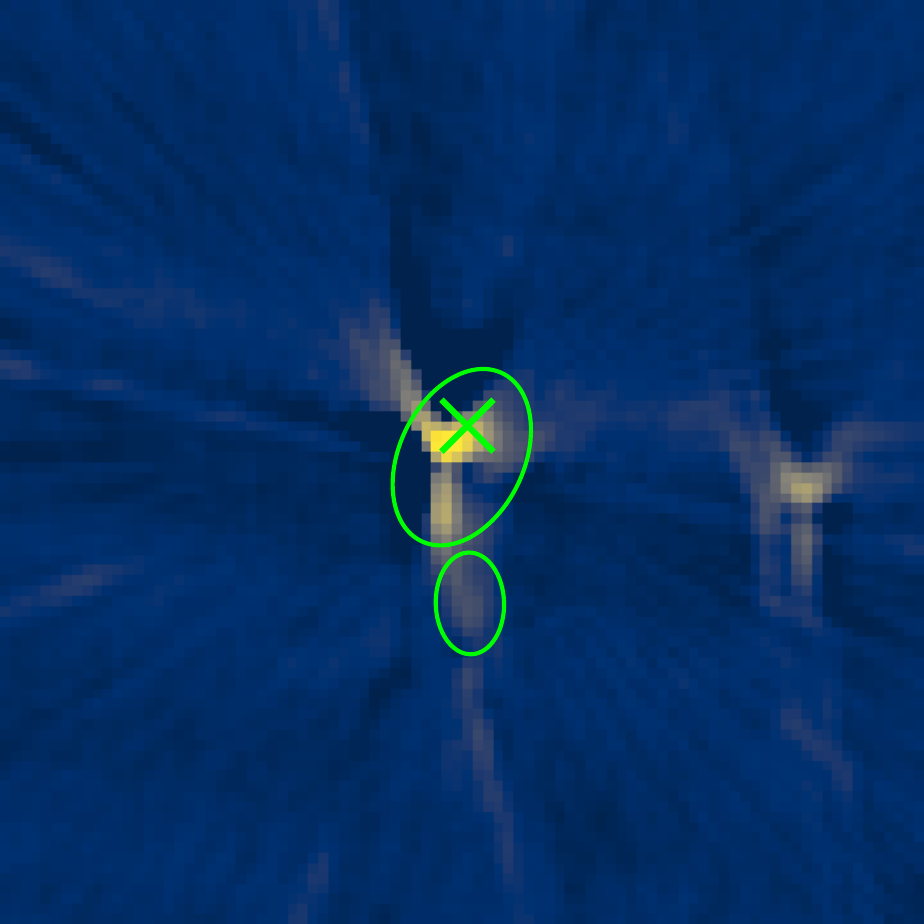}{0.33\textwidth}{(e) J221811.40-302401.1}
    \fig{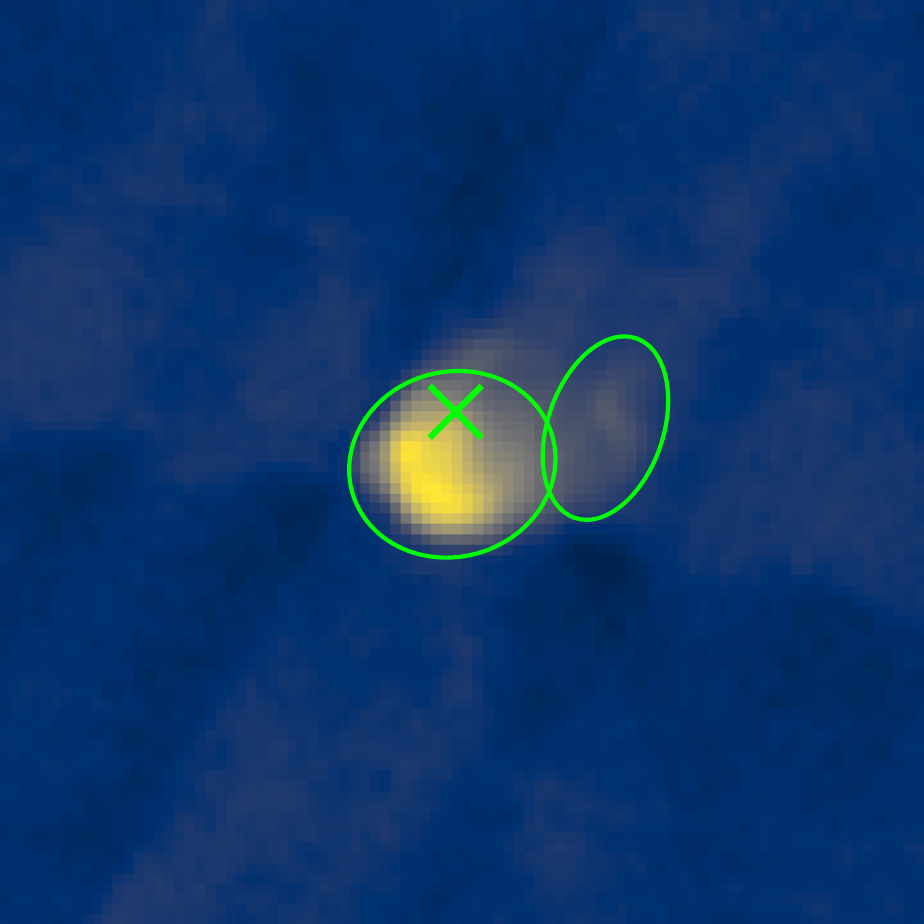}{0.33\textwidth}{(f) J184734.19-011244.2}
    }
    \caption{Images of sources which were problematic for the source-filtering approaches. (a-c) A representative sample of artifact-free sources which were identified as spurious by the Q flag, but were correctly identified as zero-artifact by our random forest model. (d-e) One-artifact sources which are correctly identified as spurious by the Q flag but were identified as having zero artifacts by our random forest classifier. These sources are representative of the types of source for which the Q flag approach works better in filtering. (f) A source which was incorrectly flagged as spurious by both the Q flag and our random forest classifier.}
    \label{fig:Q_comp}
\end{figure*}

In Fig. \ref{fig:Q_comp} we show example images of sources which were problematic for one or both of the source-filtering approaches. In particular Fig. \ref{fig:Q_comp}(a-c) show that the Q-flag is inconsistent with some artifact-free sources, including those with more interesting morphologies and extended diffuse emission. Our random forest classifier retains these less stereotypical sources more effectively than the Q flag. For some 1-artifact sources like Fig. \ref{fig:Q_comp}(d-e) the Q flag is superior in identifying the source as spurious, however the number of these sources is much less than the number of 0-artifact sources the Q flag incorrectly filters out. Finally, some sources like Fig. \ref{fig:Q_comp}(f) confused both the Q flag and random forest approaches, representing a small fraction of sources which are difficult to identify.

Our random forest model produces a more complete set of artifact-free DRAGNs than the Q flag filtering approach at the expense of a small amount of sample purity. Despite a reduction in purity, our method possesses three advantages: 1) an increase in sample completeness, including less stereotypical sources with diffuse emission, 2) an accurate estimate of the number of artifacts, and 3) the probabilities that a source contains zero, one, or two artifacts, respectively. The third advantage is the most important, as it provides an additional level of granularity not given by the Q flag method, comprehensive estimates of the probability that a source has a certain number of artifacts. These probabilities can be used to obtain a more pure set of sources, for instance by defining a probability threshold that a source has 0 artifacts and filtering out all sources that have a lower 0-artifact probability. Users can tune the completeness and purity of their samples by leveraging the artifact class probabilities provided by our random forest classifier.

%
%==================================================================================%
%==================================================================================%
%
%
%
%
%==================================================================================%
%==================================================================================%
%
\section{Discussion} \label{sec:discussion}
%==================================================================================%
\subsection{Classification Limitations}

Though the the triples-trained and log-log models attained a high F1 weighted score, there are some limitations to their classifications due to properties of their respective training sets and the VLASS DRAGNs catalog. The log-log model suffers especially in classifying 1-artifact doubles correctly, more than the triples-trained model. The greater class imbalance of the doubles dataset means that some of the more important outlier and borderline 1-artifact sources were not selected and included in the model's training set, while the triples-trained model had the luxury of including almost all the borderline values in the triples set. As such, the log-log model's ability to discriminate between these types of sources in borderline cases was less accurate, though the doubles 1-artifact class was especially more difficult to-classify compared to the analogous 2-artifact class in the triples, as discussed in Section \ref{sec:model_performance}. This could be mitigated by increasing the number of samples per grid space or altering the number of bins per axis to provide a greater chance of including borderline cases in the training set, though this means increasing the training set size and thus the amount of time and effort required to classify it.

Not all spurious detections are due to artifacts. A much smaller fraction of classifications are due to DRAGNhunter identifying multiple separate, unrelated sources as members of a single extended source. These composite sources, though less prevalent than artifact-containing sources, slightly contaminate the DRAGNs catalog with sources that often appear like regular sources but aren't real DRAGNs. These types of sources could be identified by finding potential host galaxies for each component and filtering out sources where components with large angular separation have different hosts, however this analysis is beyond the scope of this paper.

The vast majority of artifacts are sidelobes caused by bright point sources. While in many cases the bright source causing the artifacts is included in the components identified by DRAGNhunter, as is the case with most 2-artifact triples and 1-artifact doubles, all DRAGNs which contain only artifacts do not have the information about nearby sources which may be causing the artifacts. \citet{helfand_last_2015} found success using the properties of the nearest bright source to augment the data used to determine whether a source in the FIRST catalog is a sidelobe, and a similar principle could be applied to the VLASS DRAGNs to improve classification accuracy.  

There is an upper accuracy limit to the classifications of the random forest model because a small portion of sources elude clear artifact classification. This is due to difficulty in distinguishing between areas of diffuse emission from areas with elevated noise floors and is discussed in Section \ref{subsec:motiv_classification}. Sources that we had difficulty classifying were often sources where the models had difficulty making clear classifications, indicated by a source having very similar probabilities for 2 or more classes. As such, we should not expect the classifications of the model to be perfect.

%==================================================================================%
\subsection{Random Forest Hyperparameter Tuning}
Random forest models generally do not require parameter-tuning and usually provide accurate results with default settings \citep{fernandez-delgado_we_2014}. The actual performance gained by parameter-tuning is minimal \citep{probst_hyperparameters_2019}. Though we performed hyperparemter tuning using a grid search in Section \ref{sec:model_prep}, the default hyperparameters of \verb|n_estimators=100| and \verb|max_depth=None| in scikit-learn offer identical performance to the hyperparameters that our final model used. Hence, it is not necessary to tune hyperparameters to achieve optimal random forest performance. As shown in Section \ref{sec:training_sel}, training set selection methods and the size of the training set make a far larger impact on model performance.
%
%==================================================================================%
%==================================================================================%
%
%
%
%
%==================================================================================%
%==================================================================================%
%
\section{Conclusion and Future Work} \label{sec:conclusion}
Random forest models are an attractive option for identifying multi-component radio sources which contain artifacts. Due to the low computational cost, resistance to overfitting, and the need for only a relatively minimal training set when the right selection method is used, these models can be deployed quickly to increase the purity of catalogs which use noisy data. We present a catalog of artifact classifications for each source in the VLASS DRAGNs catalog found using the classifications of random forest models, where the best class is obtained from the log-log random forest model with a weighted F1 score of $97.01\%^{+1.12\%}_{-1.32\%}$, indicating high classification accuracy. Compared to the SourceFlag in column 27 of Table 6 in \citet{gordon_quick_2023} used to filter out spurious detections, our random forest method produces a more complete catalog of genuine, artifact-free DRAGNs with an estimated 99.3\% completeness and 97.7\% purity compared to 94.4\% completeness and 98.9\% purity using the traditional quantitative method in the existing catalog. Our catalog has additional benefits, as our random forest model can distinguish between the number of artifacts in a DRAGN and provide the probabilities that a source contains zero, one, or two artifacts, respectively. These probabilities can be used to create more pure sets of DRAGNs. A clean, 0-artifact subset can be derived from our catalog to create a set of VLASS DRAGNs with high completeness and purity for further astronomical analysis. 

This artifact-identification methodology using random forest models can be extended to other radio catalogs, as well as the VLASS catalog. It may be possible to use a random forest model to identify whether single sources are artifacts in the entire VLASS catalog. This is especially useful because of the high noise and elevated contamination of the existing catalog compared to other radio catalogs. A random forest model holds potential to produce a more complete subset of real sources than the current heuristic thresholds recommended in \citet{gordon_quick_2021}, as well as filter out more artifacts. 

Further, stacked, multi-epoch images from VLASS will detect more than twice as many components as detected in the \textit{Quick Look} images from individual Epochs \citep{lacy_karl_2020}. Our random forest approach will provide a robust approach identifying spurious doubles identified by DRAGNHunter on this deeper VLASS data. Beyond VLASS, our approach has the potential to benefit finding doubles in other radio surveys. For instance, \citet{ramdhanie_discovery_2023} have used DRAGNHunter with FIRST \citep{becker_first_1995} to aid in the discovery of Giant Radio Galaxies. An approach such as ours will enable DRAGNHunter to be used with FIRST to release a full catalog of double and triple radio sources found by the survey. Moreover much deeper radio surveys, such as EMU are expected to find hundreds of thousands of DRAGNs \citep{norris_emu_2025}, necessitating automated detection algorithms. The approach presented in this paper has the potential to improve the efficiency of automated double finding algorithms employed by ultra data rich radio surveys such as EMU.

%% Please use the acknowledgment and contribution environments. This will 
%% be anonomyized when the "anonymous" style option is used. 
\begin{acknowledgments}
The authors thank Thane Goetz and Mia Benedetto for assisting with the visual classification of the triples as part of another project. V.E. is supported by the Doherty Research Fellowship through the University of Wisconsin-Madison.

The National Radio Astronomy Observatory is a facility of the National Science Foundation operated under cooperative agreement by Associated Universities, Inc. CIRADA is funded by a grant from the Canada Foundation for Innovation 2017 Innovation Fund (Project 35999), as well as by the Provinces of Ontario, British Columbia, Alberta, Manitoba and Quebec. 

This publication makes use of data products from the Wide-field Infrared Survey Explorer, which is a joint project of the University of California, Los Angeles, and the Jet Propulsion Laboratory/California Institute of Technology, funded by the National Aeronautics and Space Administration.
\end{acknowledgments}

\begin{contribution}
V.E. led the investigation, formal analysis and writing of the manuscript. E.H. and M.M. assisted with visual classification, interpretation, and manuscript review. S.B. assisted with visual classification. Y.G. assisted with interpretation and manuscript review.  

\end{contribution}

\facilities{VLA, WISE}

\software{
    scikit-learn \citep{pedregosa_scikit-learn_2011}, 
    NumPy \citep{harris_array_2020}, 
    Pandas \citep{team_pandas-devpandas_2025}, 
    Matplotlib \citep{hunter_matplotlib_2007}, 
    Seaborn \citep{waskom_seaborn_2021}, 
    AstroPy \citep{2013A&A...558A..33A,2018AJ....156..123A,2022ApJ...935..167A}
    SAOImage DS9 \citep{payne_new_2003}
          }

\appendix

\section{Catalog of Random Forest Classification Results} \label{appendix:catalog}

We provide a machine-readable catalog of the best estimate classification, as well as the individual class-wise probabilities from both the triples-trained and the log-log doubles-selected models. We recommend using the best\_class as this gives the artifact classification from the log-log LAS S/N and flux S/N doubles model which has a lower rate of misclassifying 0-artifact sources as containing artifacts. Filtering Table \ref{tab:catalog} for sources where best\_class == 0 and cross-referencing these sources with their corresponding entries in table 6 of \citet{gordon_quick_2023} will yield a catalog comprised of approximately 97.7\% artifact-free doubles sources. 

The class-wise probabilities as obtained from both models can be used to obtain more pure subsets of the DRAGNs catalog, though this will also result in a loss of completeness of artifact-free sources. The triples-trained model generally assigns a higher probability for a given source to contain artifacts and as such will give a more pure catalog. Conversely, the log-log model is slightly more conservative in artifact probabilities due to properties of doubles discussed in section \ref{sec:model_performance}. The visual artifact classifications of the triple sources will be available in future work as part of more intensive visual classification addressing source morphology and the accuracy of the components DRAGNhunter uses.

\begin{deluxetable}{c c c c c c c c c}[htb!]
    \tablecaption{Example entries from the accompanying VLASS DRAGNs artifact catalog}
    \label{tab:catalog}
    \tablehead{
        \colhead{Name\tablenotemark{a}} & \colhead{Best\_class\tablenotemark{b}} & \colhead{Visual\_class\tablenotemark{c}} & \colhead{Trip\_prob\_0} & \colhead{Trip\_prob\_1} & \colhead{Trip\_prob\_2} & \colhead{Loglog\_prob\_0} & \colhead{Loglog\_prob\_1} & \colhead{Loglog\_prob\_2}
    }    
    \startdata
    J000002.90+095706.5 & 0 & -- & 0.9925 & 0.0075 & 0.0 & 0.995 & 0.0025 & 0.0025 \\
    J000010.00+792237.1 & 0 & -- & 0.965 & 0.02 & 0.015 & 0.965 & 0.01 & 0.025 \\
    J000012.10+291114.9 & 0 & -- & 0.9875 & 0.01 & 0.0025 & 0.905 & 0.0375 & 0.0575 \\
    J000019.36-272515.1 & 0 & 0.0 & 0.965 & 0.0225 & 0.0125 & 0.98 & 0.0125 & 0.0075 \\
    J000020.63-322117.7 & 2 & -- & 0.2151 & 0.1325 & 0.6524 & 0.1225 & 0.0025 & 0.875 \\
    J000027.13-331948.3 & 0 & -- & 0.9925 & 0.0025 & 0.005 & 0.965 & 0.035 & 0.0 \\
    J000031.36-192157.7 & 0 & -- & 0.9875 & 0.0075 & 0.005 & 0.9975 & 0.0 & 0.0025 \\
    J000039.66+041127.3 & 0 & -- & 0.9875 & 0.01 & 0.0025 & 0.94 & 0.06 & 0.0 \\
    J000040.21-142346.2 & 0 & -- & 0.9975 & 0.0025 & 0.0 & 1.0 & 0.0 & 0.0 \\
    J000042.17-342400.4 & 0 & -- & 0.99 & 0.0075 & 0.0025 & 1.0 & 0.0 & 0.0 \\
    % ... & ... & ... & ... & ... & ... & ... & ... & ...
    \enddata

    \tablenotetext{a}{J2000 name of the VLASS DRAGN from table 6 of \citet{gordon_quick_2023}.}
    \tablenotetext{b}{Best random forest artifact classification of the source as taken from the log-log model.}
    \tablenotetext{c}{Visual classifications are only available for sources that were part of either the doubles verification set or the log-log selected doubles training set, 804 total.}
    % \tablenotetext{d}{Probability that the source contains 0 artifacts as obtained from the log-log model.}
    % \tablenotetext{e}{Probability that the source contains 0 artifacts as obtained from the triples model.}
\end{deluxetable}

%% For this sample we use BibTeX plus aasjournalv7.bst to generate the
%% the bibliography. The sample7.bib file was populated from ADS. To
%% get the citations to show in the compiled file do the following:
%%
%% pdflatex sample7.tex
%% bibtext sample7
%% pdflatex sample7.tex
%% pdflatex sample7.tex

\bibliography{references}{}
\bibliographystyle{aasjournalv7}

%% This command is needed to show the entire author+affiliation list when
%% the collaboration and author truncation commands are used.  It has to
%% go at the end of the manuscript.
%\allauthors

%% Include this line if you are using the \added, \replaced, \deleted
%% commands to see a summary list of all changes at the end of the article.
%\listofchanges

\end{document}